\documentclass[trackchanges,twocolumn,twocolappendix]{aastex7}

\usepackage[utf8]{inputenc} 
\usepackage[T1]{fontenc} 
\usepackage{graphicx}
\usepackage{amsmath}
\usepackage{amsfonts} 
\usepackage{tabularx}
\usepackage{booktabs}
\usepackage{bm}
\usepackage{url}
\usepackage{natbib}
\usepackage{hyperref}
\usepackage{nameref}

\begin{document}

\title[Probing the LSS with 2PACF and power spectrum]{Probing large-scale structures with the two-point function and the power spectrum: \\
insights into cosmic clustering evolution}

\author[orcid=0000-0002-6320-425X,gname='Camila',sname='Franco']{Camila Franco}
\affiliation{Observatório Nacional, Rua General José Cristino, 77, São Cristóvão, 20921-400, Rio de Janeiro, RJ, Brazil}
\email[show]{camilafranco@on.br}  

\author[orcid=0000-0002-0562-2541,gname=Felipe, sname='Avila']{Felipe Avila} 
\affiliation{Observatório Nacional, Rua General José Cristino, 77, São Cristóvão, 20921-400, Rio de Janeiro, RJ, Brazil}
\email{fsavila2@gmail.com}

\author[orcid=0000-0003-3034-0762,gname=Armando,sname=Bernui]{Armando Bernui}
\affiliation{Observatório Nacional, Rua General José Cristino, 77, São Cristóvão, 20921-400, Rio de Janeiro, RJ, Brazil}
\email{bernui@on.br}

\begin{abstract}
Understanding the large-scale structure of the Universe requires analyses of cosmic clustering and its evolution over time. 
In this work, we investigate the clustering properties of SDSS blue galaxies, which are excellent tracers of dark matter, along two distinct epochs of the Universe, utilizing estimators like the two-point angular correlation function (2PACF), the angular power spectra, among others. 
Considering a model-independent approach, we perform analyses in two disjoint redshift shells, $0 \leq z < 0.06$ and $0.06 \leq z < 0.12$, to investigate the distribution of large cosmic structures. 
Using Bayesian inference methods, we constrain the parameter 
that quantifies the galaxy clustering in the 2PACF, enabling us to perform comparisons among different regions on the sky and between different epochs in the Universe regarding the gravitational action on matter structures. 
Our analyses complement previous efforts to map large-scale structures in the Local Universe. 
In addition, this study reveals differences regarding the clustering of large cosmic structures comparing two epochs of the Universe, analyses done with diverse estimators. 
Results reveal, clearly, distinct evolutionary signatures between the two redshift shells. 
Moreover, we had the opportunity to test the concordance cosmological model under extreme conditions in the highly non-linear Local Universe, computing the amplitude of the angular power spectrum at very small scales. 
Ultimately, all our analyses serve as a set of consistency tests of the concordance cosmological model, the $\Lambda$CDM.
\end{abstract}



\section{Introduction}
\label{sec:intro}

The Universe is plenty of complex large structures whose features are unnoticed, in principle, just by observing the mapped cosmic objects in the sky (2-dimensional, 2D, projected data) or its distribution in redshift space~\citep{Valade,Hoffman}. 
The most interesting features come from a collective evolutionary phenomenon that is predominant throughout the history of the Universe: matter clustering. 
Revealing it means describing, at the same time, the growth of 
cosmic structures (from primordial over-densities) and the growth of voids (from primordial under-densities). 
With the advent of large and deep astronomical surveys, such structures in the Local Universe are now being revealed and studied~\citep{Courtois12,Courtois13,Courtois25,Hoffman,Lopes24}, although they were predicted in cosmological simulations two decades ago~\citep{Springel05,Schaye15,Pillepich18}. 


In fact, the observed distribution of galaxies in deep surveys presents an intricate network of structures, like filaments, walls, clusters, and voids, forming what has been called the cosmic web, whose description provides crucial insights into the structure and dynamics of the Universe. 
Filaments are large-scale, thread-like structures in the cosmic web, formed by concentrations of galaxies and dark matter~\citep{Bond96,Sousbie11,Sarkar25,Holm25}; 
walls are vast, sheet-like structures in the cosmic web made up 
superclusters of galaxies and dark matter~\citep{Ramella92,Gavazzi10,Einasto11}; 
clusters and superclusters are dense groupings of galaxies held together by gravity, often containing hundreds to thousands of galaxies, as well as hot gas and dark matter~\citep{Gunn72,Press74,White91,Springel05}; 
voids, instead, are vast and empty regions in the Universe with very few galaxies or matter~\citep{Sheth04,Tully08,Pan12}.

Tomographic analyses in redshift bins can help to comprehend the clustering evolution of cosmic structures~\citep{Budavari03,papovich08,Sawangwit11,Asorey12,Donoso14,Marques20}. 
For this, we study clustering properties of galaxies on two disjoint redshift bins using model-independent 
statistical tools. 
Our main objectives are: 
(i)~to complement the analyses reported in the literature mapping the cosmic structures at large scales in the Local Universe, revealing large over-dense regions with highly clustered matter and under-dense regions almost void of galaxies~\citep{Courtois12,Kitaura12,Pomarede13,Cybulski14,Nuza14}; 
(ii)~to reveal differences regarding the clustering of large cosmic structures comparing two, close but different, epochs in the Universe evolution~\citep{Marques20,Franco2024}; 
(iii)~to test the concordance cosmological model under extreme conditions in the highly non-linear Local Universe, $z \simeq 0$, quantifying the amplitude of the angular power spectrum at very small scales~\citep{Wu25,Franco2025}; 
and, 
(iv)~to perform statistical isotropy examination of the Local Universe in both redshift bins~\citep{Alonso15,Novaes2018}.

This work is organized as follows: Section~\ref{sec:data} describes the observational data used in our analyses. In Section~\ref{sec:method}, we outline the theoretical framework of the 2PACF, the covariance matrix, the angular power spectrum, and the role of the parameters involved 
in this study, along with the methodology for Bayesian parameter inference. The results and discussions 
are presented in Section~\ref{sec:results}. 
Finally, conclusions are provided in Section~\ref{sec:conclusions}.

\section{Sloan Digital Sky Survey data}
\label{sec:data}

\begin{figure}
    \begin{minipage}[b]{\linewidth}
        \centering
        \includegraphics[width=0.9\textwidth]{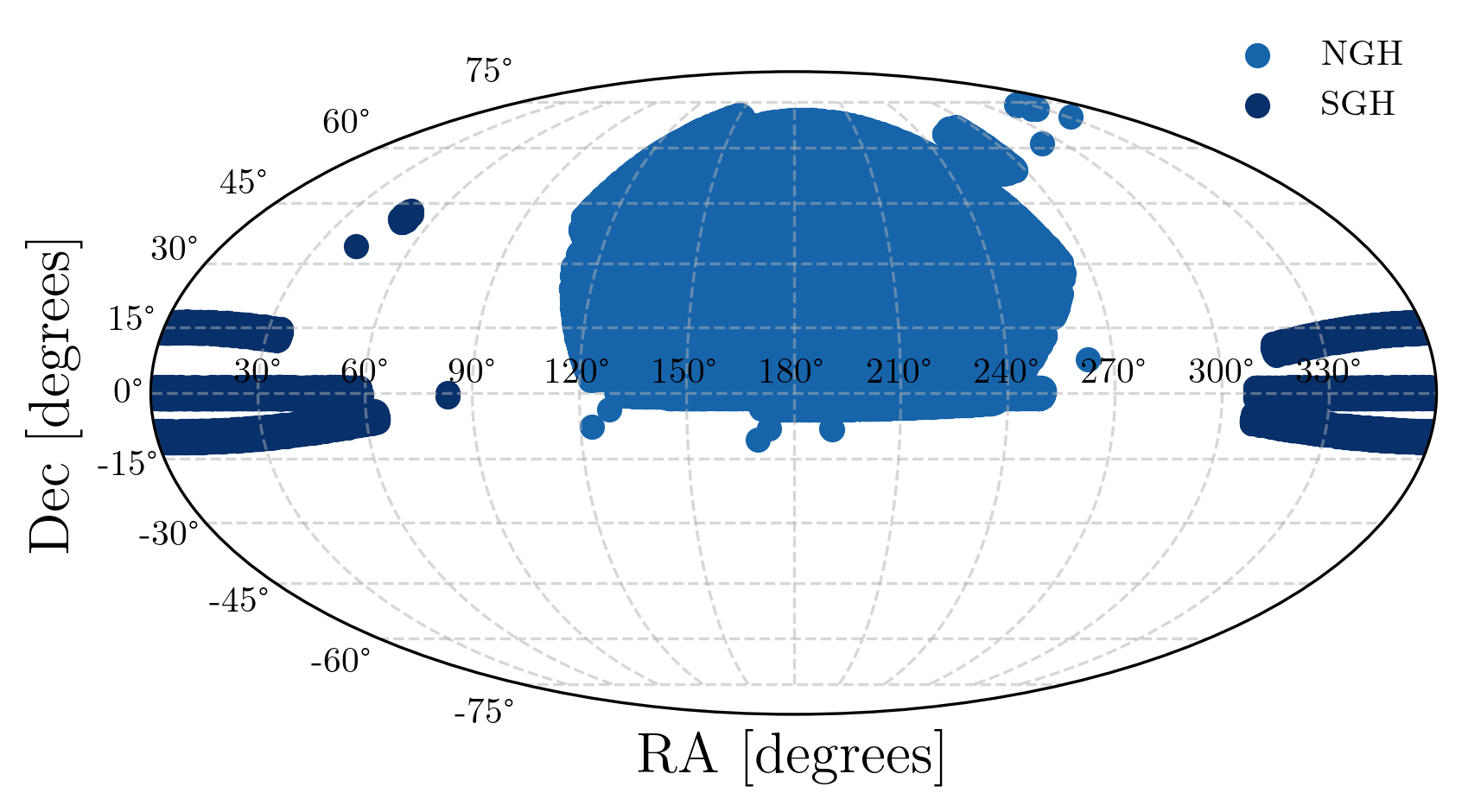}
    \end{minipage}
    \caption{SDSS footprint in equatorial coordinates. The Northern Galactic Hemisphere (NGH), shown in blue, is the central region of the projection, while the Southern Galactic Hemisphere (SGH), in dark blue, is located at the edges. Both regions have irregular boundaries, with the NGH exhibiting higher density of observations compared to the SGH.}
    \label{fig:sdss_footprint}
\end{figure}

Our analyses are carried out using star-forming blue galaxies from the Sloan Digital Sky Survey~\citep[SDSS; ][]{York2000} selected through the color-color diagram, following the procedure described in~\cite{Avila2019}. 
The objects were made available in the twelfth data release (DR12) of SDSS~\citep{Alam2015}; 
the sky footprint of these data is shown in Figure~\ref{fig:sdss_footprint}. 

Blue galaxies are predominantly spiral and are observed also in low-density environments, such as cosmic voids~\citep{Schneider2006,Hoyle2012}. These galaxies are characterized by active star-formation, which give rise to their distinct blue color, primarily due to the presence of young, massive stars~\citep{Mo2010}. They are also less clustered than their red counterparts, 
a cosmic feature manifested in its bias relative to matter close to 1, i.e., $b \approx 1$, 
highlighting their distinct evolutionary processes in comparison to galaxies in high-density regions~\citep{Hoyle2012,Dressler1980,Postman1984,Strateva2001}. 
All these features make blue galaxies highly interesting objects for the type of analyses we aim to perform and also allow us to compare our results with previous studies, such as those using extragalactic HI sources~\citep{Franco2024,Wu25}.

\begin{figure}
    \begin{minipage}[b]{\linewidth}
        \centering
        \includegraphics[width=0.9\textwidth]{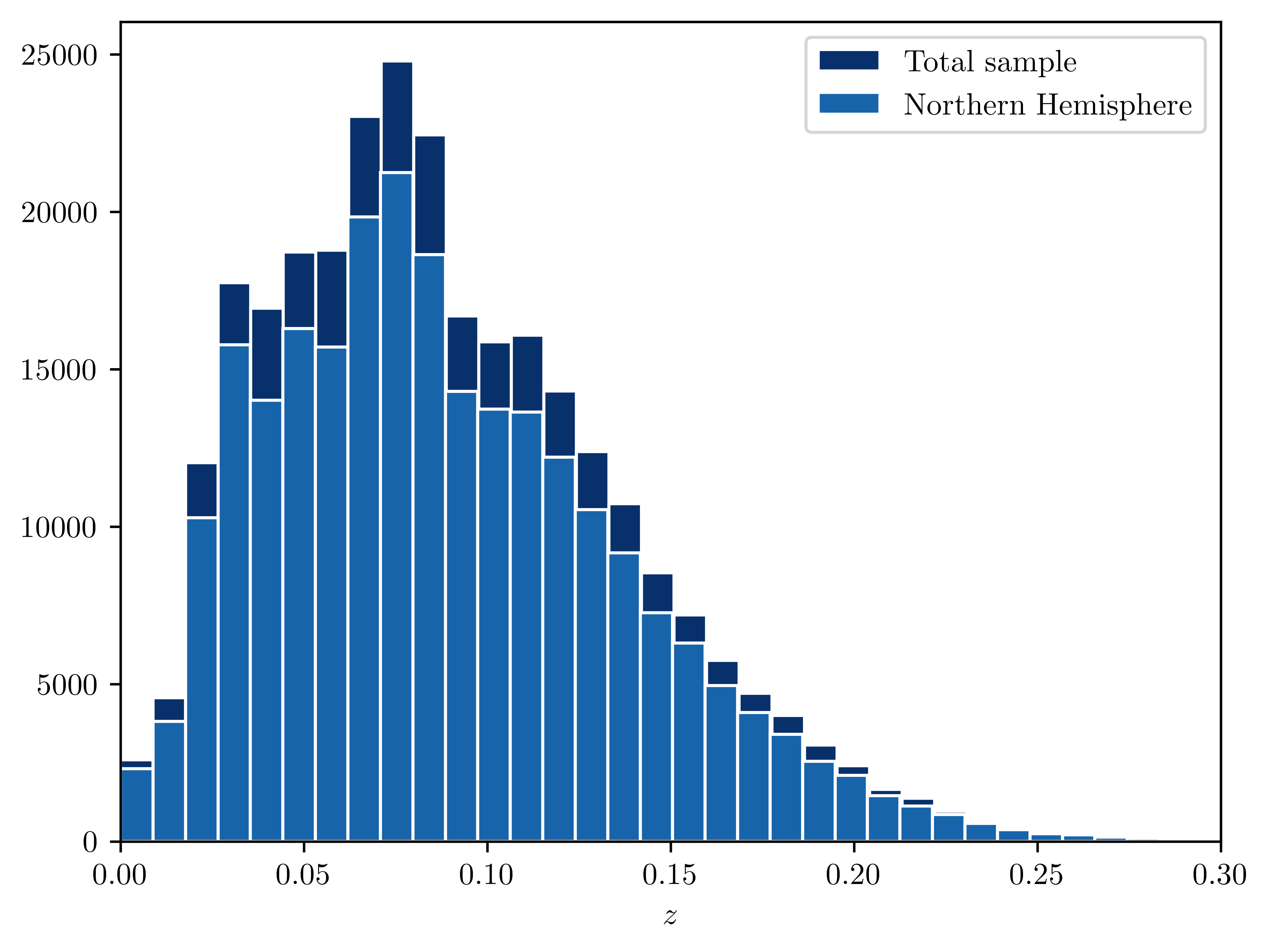}
    \end{minipage}
    \caption{Redshift distribution of the SDSS blue galaxies selected sample compared to the total sample.}
    \label{fig:sdss-hist}
\end{figure}
\begin{figure}
\begin{minipage}[b]{\linewidth}
\centering
\includegraphics[width=0.9\textwidth]{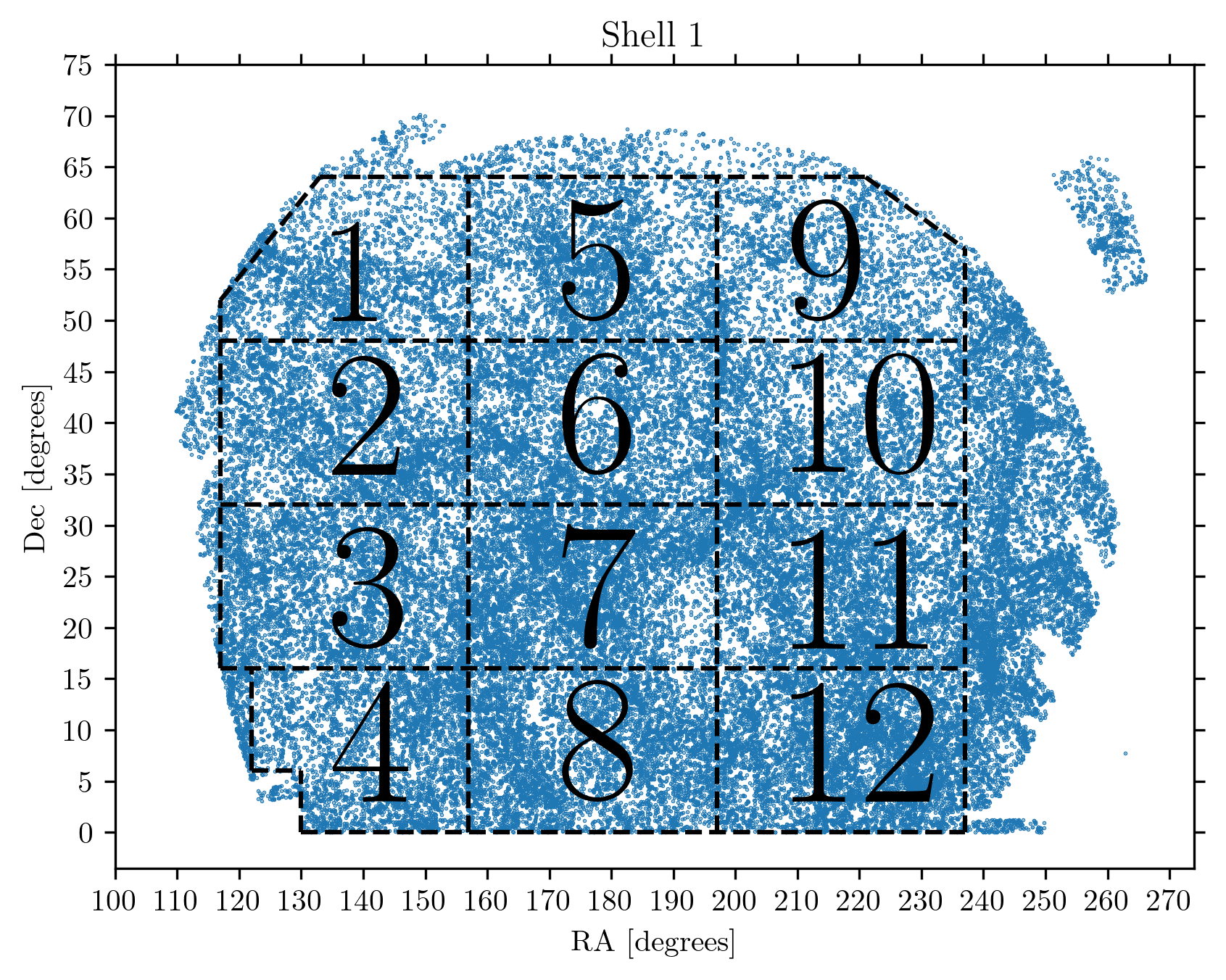}
\includegraphics[width=0.9\textwidth]{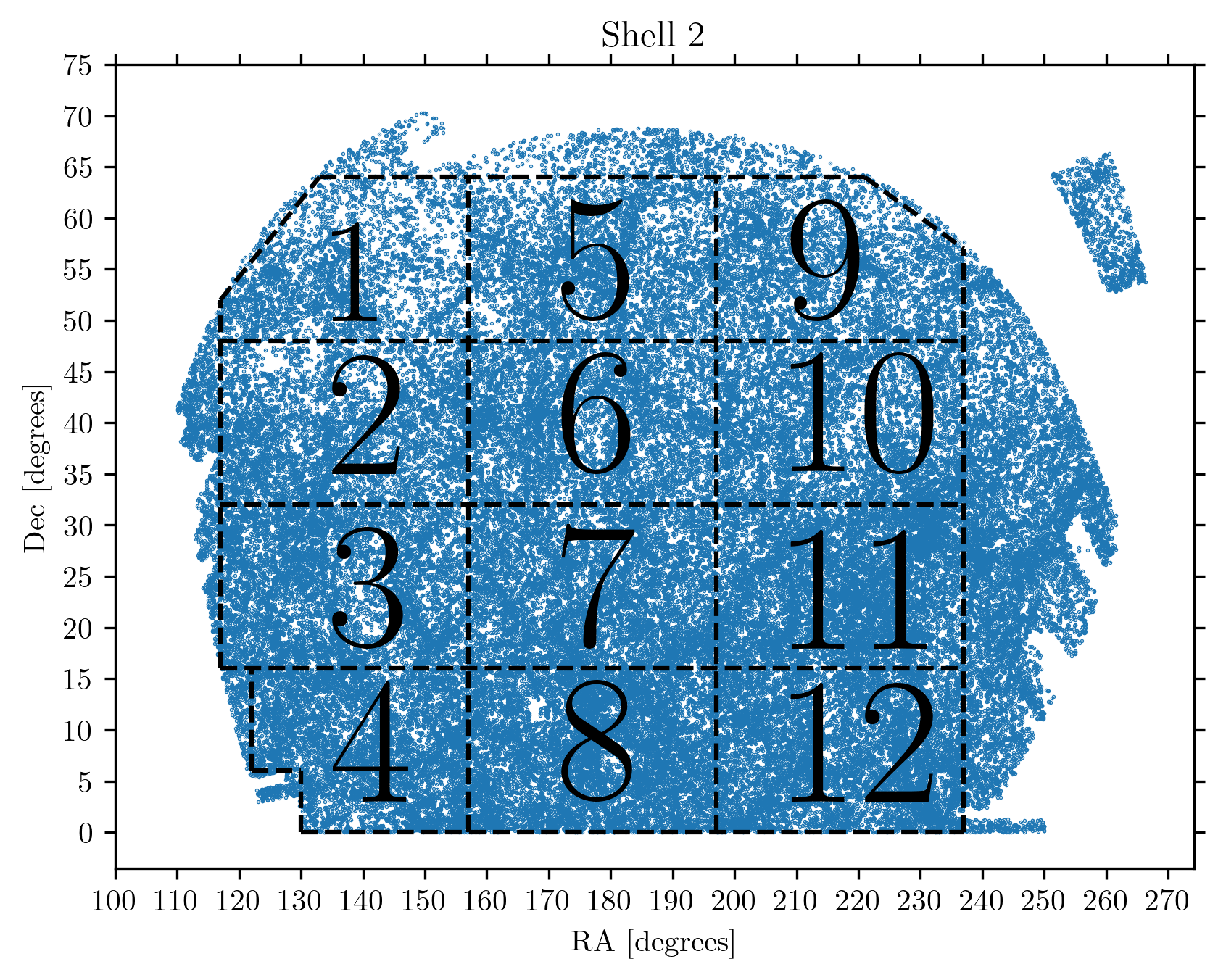}
\end{minipage}
\caption{Footprint of the selected sample, located in the NGH. We divided each shell into $12$ Areas. \textbf{Upper panel}: Shell 1 ($0 \leq z < 0.06$) contains $62,495$ galaxies; \textbf{Bottom panel}: Shell 2 ($0.06 \leq z < 0.12$) contains $96,712$ galaxies.}
\label{fig:footprint_shells}
\end{figure}

For our directional analyses, many disjoint regions as possible are needed, with comparable areas and number densities and in accordance with the lower limit of the angular homogeneity scale, $\theta_{H} \simeq 20^{\circ}$~\citep{Avila18,Avila2019}. To meet these criteria while preserving the statistical reliability, we selected the Northern Galactic Hemisphere (NGH) and subdivided our sample into $12$ regions, each with angular dimensions $\Delta\text{RA}\simeq 40^{\circ}$ and $\Delta\text{Dec}\simeq 16^{\circ}$, within angular coordinates $117^{\circ} \leq \text{RA} < 237^{\circ}$ and $0^{\circ} \leq \text{Dec} < 64^{\circ}$. 
Figure~\ref{fig:sdss-hist} shows the redshift distribution of our sample compared to the full sample, and Figure~\ref{fig:footprint_shells} displays the division into the 12 Areas for our analysis.

This configuration was chosen based on two main considerations: 
(i) preserving the angular scale of homogeneity in the selected regions, and 
(ii) maximizing the number of regions for a meaningful comparative directional analysis. 
After testing several configurations, the chosen dimensions provided an optimal balance between spatial resolution and statistical robustness. 


Furthermore, the shape of the regions was determined in order to facilitate the replication of the footprint in the mocks and random catalogs. To this end, only galaxies located within the regular polygonal boundary (see Figure~\ref{fig:footprint_shells}) were included in the final analysis.

Moreover, in this work we aim to perform model-independent analyses which means that we shall focus on projected 2D data, and in 3D study using the Hubble-Lama\^{\i}tre law to calculate physical distances with cosmography. For this, we select the SDSS blue galaxies within the redshift $0 \leq z < 0.12$. 
We perform tomographic analyses in two thin shells, termed 
Shell 1 ($0 \leq z < 0.06$) and 
Shell 2 ($0.06 \leq z < 0.12$). 
The width of the shells is $\delta z = 0.06$. 
For each shell, the galaxies are projected on the celestial sphere, as seen in Figure~\ref{fig:footprint_shells}. 
The observational features of each region within each shell are listed in Table~\ref{tab:features-areas}. 
In the end, under these conditions, our selected sample contains $159,207$ blue galaxies: 
$62,495$ galaxies in Shell 1 and 
$96,712$ galaxies in Shell 2. 

%

\begin{table*}[!ht]
\centering
\caption{Features of the 12 regions of the 
Shell 1 ($0 \leq z < 0.06$) and 
Shell 2 ($0.06 \leq z < 0.12$). 
The uncertainties in the number density, $n$, were calculated using $\sigma_{n_i} = \sqrt{N_i}/\text{area}_i$, where $N_i$ is the number of blue galaxies in $\text{area}_i$. The total volumes of Shells 1 and 2 are, respectively, $V_1 = 1.1\times 10^{-4}\text{Gpc}^{3}$ and $V_2 = 7.3\times 10^{-3}\text{Gpc}^{3}$.
} 
\begin{tabularx}{\linewidth}{>{\centering\arraybackslash}X >{\centering\arraybackslash}X >{\centering\arraybackslash}X >{\centering\arraybackslash}X >{\centering\arraybackslash}X >{\centering\arraybackslash}X }
\hline
\hline
& & \multicolumn{2}{c}{Shell 1} & \multicolumn{2}{c}{Shell 2}\\
\cmidrule(lr){3-4} \cmidrule(lr){5-6} 
& area [deg$^2$] & galaxies & {$n\,[\text{deg}^{-2}]$} & galaxies & {$n\,[\text{deg}^{-2}]$}\\
\midrule
Area 1 & $544$ & $3,154$ & $5.80\,\pm\,0.10$ & $3,924$ & $7.21\,\pm\,0.12$\\
Area 2 & $640$ & $4,975$ & $7.77\,\pm\,0.11$ & $6,515$ & $10.18\,\pm\,0.13$\\
Area 3 & $640$ & $5,611$ & $8.77\,\pm\,0.12$ & $8,614$ & $13.46\,\pm\,0.15$\\
Area 4 & $512$ & $4,292$ & $8.38\,\pm\,0.13$ & $8,171$ & $15.96\,\pm\,0.18$\\
Area 5 & $640$ & $4,292$ & $6.71\,\pm\,0.10$ & $5,921$ & $9.25\,\pm\,0.12$\\
Area 6 & $640$ & $5,272$ & $8.24\,\pm\,0.11$ & $7,426$ & $11.60\,\pm\,0.13$\\
Area 7 & $640$ & $7,233$ & $11.30\,\pm\,0.13$ & $9,029$ & $14.11\,\pm\,0.15$\\
Area 8 & $640$ & $6,170$ & $9.64\,\pm\,0.12$ & $12,144$ & $18.98\,\pm\,0.17$\\
Area 9 & $584$ & $2,736$ & $4.68\,\pm\,0.09$ & $5,403$ & $9.25\,\pm\,0.13$\\
Area 10 & $640$ & $4,390$ & $6.86\,\pm\,0.10$ & $7,088$ & $11.08\,\pm\,0.13$\\
Area 11 & $640$ & $6,342$ & $9.91\,\pm\,0.12$ & $11,223$ & $17.54\,\pm\,0.17$\\
Area 12 & $640$ & $8,028$ & $12.54\,\pm\,0.14$ & $11,254$ & $17.58\,\pm\,0.17$\\
\bottomrule
\end{tabularx}
\label{tab:features-areas}
\end{table*}

\section{Methodology}
\label{sec:method}

In this section, we outline the methodology used to study the distribution of blue galaxies in the SDSS data. It is worth mentioning that our study is based on the angular positions of the galaxies and in their spectroscopic redshifts, that one can use when applying cosmography, ensuring a model-independent approach. 
However, to stablish the significance of our results one needs to perform comparisons with the outcomes expected in the concordance cosmological model, $\Lambda$CDM, and for this we use a large set of mocks, data produced assuming a fiducial cosmology.

\subsection{Two-point angular correlation function (2PACF)}
\label{sec:tpacf}
The spatial distribution of objects can be effectively characterized using the two-point angular correlation function (2PACF), that has become a cornerstone in observational cosmology for analysing large-scale structures. This powerful statistical tool quantifies the excess probability of finding two objects separated by a given angular distance $\theta$ compared to a random distribution. Several estimators have been proposed in the literature for measuring the 2PACF~\citep{Hewett82,DP83,Hamilton93,PH74}. Among these, the Landy-Szalay~\citep[LS; ][]{LS93} estimator has proven to be the most robust for count-count correlations, and is defined as
\begin{equation}
    \label{eq:ls}
    \omega(\theta) = \frac{DD(\theta) - 2DR(\theta) + RR(\theta)}{RR(\theta)},
\end{equation}
where $DD(\theta)$ is the number of galaxy pairs in the sample data with angular separation $\theta$, normalized by the total number of pairs; $RR(\theta)$ is a similar quantity, but for the pairs in a random sample; and $DR(\theta)$ corresponds to a cross-correlation between a data object and a random object. The angular separation between pairs is calculated using the trigonometric relationship
\begin{equation}
\label{eq:theta}
    \theta_{ij} = \cos^{-1}[\sin(\delta_i)\sin(\delta_j) + \cos(\delta_i)\cos(\delta_j)\cos(\alpha_i - \alpha_j)] \,,
\end{equation}
where $\alpha_i$, $\alpha_j$ and $\delta_i$, $\delta_j$ are the right ascension and the declination, respectively, of the galaxies $i$ and $j$.

To measure the 2PACF, we used the public code \textsc{treecorr}\footnote{\url{https://rmjarvis.github.io/TreeCorr/_build/html/index.html}}~\citep{Jarvis2015}, applying $22$ linearly spaced bins within the angular range $[\theta_{\text{min}}, \theta_{\text{max}}] = [0.001^{\circ}, 25^{\circ}]$ for the large-angle scenario. Considering the small-angle scenario, we used $12$ logarithmically spaced bins within the angular range $[\theta_{\text{min}}, \theta_{\text{max}}] = [0.001^{\circ}, 10^{\circ}]$. A random catalog was constructed for this purpose, maintaining the same angular footprint as the original region but with a uniform distribution of points and a number density $10$ times greater than that of the observed dataset. For further details on random catalog construction, see, e.g., \cite{deCarvalho18, Keihanen19, Wang13}.

\subsection{Log-normal simulations}
\label{sec:mocks}

The covariance matrix was used to estimate uncertainties, as it provides reliable error estimates by accounting for correlations between data points. 

To compute the covariance matrix, we adopted log-normal random field distributions~\citep{Coles91}, which have been widely used in similar studies~\citep{Franco2024,Avila24}. The mock catalogs were generated with the public code developed by \cite{Agrawal17}\footnote{\url{https://bitbucket.org/komatsu5147/lognormal_galaxies/src/master/}}, which produces galaxy distributions that statistically match the large scale structure of the Universe under a $\Lambda$CDM cosmology consistent with current observational constraints by ~\cite{Planck20}, i.e., considering the fiducial cosmological parameters presented in Table~\ref{tab:input-cosmo}.

\begin{table}[!ht]
\centering
\caption{Fiducial cosmological parameters from~\citet{Planck20}.}
\begin{tabularx}{0.8\linewidth}{>{\centering\arraybackslash}X}
\hline
\hline
Cosmological parameters\\
\hline
$\Omega_b h^2 = 0.02236$\\
$\Omega_c h^2 = 0.1202$\\
$\ln(10A_s) = 3.045$\\
$n_s = 0.9649$\\
$\Sigma m_{\nu} = 0.06$ eV\\
$h = 0.6727$\\
\hline
\end{tabularx}
\label{tab:input-cosmo}
\end{table}

The other input parameters required to create these catalogs are the redshift $z$, the bias $b$, the number of galaxies $N_{g}$, and the box dimensions ($L_{x}, L_{y}, L_{z}$), detailed in Table~\ref{tab:input-mock}. 
Following this approach, $N = 1,000$ mock catalogs were generated to ensure statistical reliability. 
For further details on the pipeline to produce the set of mocks see, e.g.,~\cite{Franco2024}. 

\begin{table}[!ht]
\centering
\caption{Survey configuration used to generate the set of $1,000$ log-normal mock catalogs used in our analyses. $N_g$ is the number of galaxies in the box configuration, with dimensions $(L_x,L_y,L_z)$; $z$ is the redshift; and $b$ is the bias~\citep{Avila24}.}
\begin{tabularx}{0.8\linewidth}{>{\centering\arraybackslash}X}
\hline
\hline
Survey configuration\\
\hline
$z = 0.08$\\
$b = 1.1$\\
$N_g = 5 \times 10^6$ \\
$L_x = 750$  Mpc $h^{-1}$ \\
$L_y = 1,200$ Mpc $h^{-1}$ \\
$L_z = 750$  Mpc $h^{-1}$ \\
\hline
\end{tabularx}
\label{tab:input-mock}
\end{table}

The covariance matrix was calculated using the following expression,
\begin{equation}\label{eq:cov}
    \text{Cov}_{ij} = \frac{1}{N} \sum_{k = 1}^{N}[\omega_k(\theta_i) - \overline{\omega}_k(\theta_i)][\omega_k(\theta_j) - \overline{\omega}_k(\theta_j)],
\end{equation}
where the indices $i, j = 1,2,...,N_b$ represent each bin $\theta_i$; $\omega_k$ is the 2PACF for the $k$-th mock ($k = 1,2,...,N$); $\overline{\omega}(\theta_i)$ and $\overline{\omega}(\theta_j)$ are the mean value at the bin $i$ and $j$, respectively. This approach allows for a more precise quantification of uncertainties, ensuring the robustness of the results presented in this work. 

\subsection{Angular Power Spectrum}
\label{sec:ang_spectrum}

Matter fluctuations can be described using the two-point correlation function, either in Fourier space through the power spectrum, $P(k)$, where $k$ is the magnitude of the wave vector, or in real space~(configuration space) via the correlation function, $\xi(r)$, where $r$ represents the physical separation between two objects~\citep{Peebles80}. These two approaches are Fourier transform pairs, with $k$ and $r$ being inversely proportional. 
The choice between these methods for studying large-scale structures depends on the specific goals of the analysis, as each may be more suitable for different applications. Currently, in collaborations such as DESI, both estimators $\xi(r)$ and $P(k)$ are often presented to provide complementary insights~\citep{DESI24}.

Both two-point correlation functions, $P(k)$ and $\xi(r)$, can be studied from data projected onto the celestial sphere. 
In the previous section, we saw the angular correlation function, $\omega(\theta)$, which is the angular or projected version of    $\xi(r)$. 
For the projected version of $P(k)$, we first need to define the matter fluctuation in the sphere, at a given epoch $z_{i}$, 
\begin{equation}\label{eq:number_fluctuation}
\delta_i(\theta,\phi) \equiv \frac{\rho_i(\theta,\phi)-\bar{\rho}_i}{\bar{\rho}_i},
\end{equation}
where $\bar{\rho}_i$ is the average number density over the sky at the epoch $z_{i}$. 
Note that we avoid using distances so as not to use a cosmological model in their determination. 
%
Then, having defined the fluctuation of matter in the celestial sphere, the decomposition into spherical harmonics is applied~(for the theoretical methodology and its applications, see, e.g.,~\cite{Peebles80,Tegmark02,Thomas11,Leistedt13,Ando18,Fang20}),
\begin{equation}\label{eq:spherical_harmonics}
\delta_i(\theta, \phi) = \sum_{\ell=0}^{\infty} \sum_{m=-\ell}^{\ell} a_{\ell,m} Y_{\ell,m}(\theta, \phi),  
\end{equation}
where the power spectrum, $C_{\ell}$, is obtained from the variance of the coefficients $a_{\ell,m}$ 
\begin{equation}\label{eq:Cl_full}
C_{\ell} = \langle |a_{\ell,m}|^{2}\rangle \,,
\end{equation}
and the multipole moment, $\ell$, is related to the angular scale, $\theta$, 
by $\ell = 180^{\circ}/\theta$, with $\theta$ measured in degrees. 
Since we will be using the public code \textsc{ccl}\footnote{\url{https://github.com/LSSTDESC/CCL}}~\citep{Chisari19} to obtain the theoretical curve of $C_{\ell}$, we will use their notation to define $C_{\ell}$ in function of the matter power spectrum. For two tracers, $a$ and $b$, the angular power spectrum can be written as
\begin{equation}\label{eq:curva_teorica}
    C_{\ell}^{ab} = 4\pi \int_{0}^{\infty} \frac{dk}{k}\mathcal{P}_{\Phi}(k) \Delta_{\ell}^{a}(k) \Delta_{\ell}^{b}(k),
\end{equation}
where $\mathcal{P}_{\Phi}(k)$ is the dimensionless power spectrum of the primordial curvature perturbations, and $\Delta_{\ell}^{a}(k)$ and $\Delta_{\ell}^{b}(k)$ are the transfer functions corresponding to these tracers. 
In our case, we are dealing with discrete sources, these transfer functions can be calculated as
\begin{equation}
    \Delta_{\ell}(k) = \int dz\, p_z(z)b(z)T_{\delta}(k,z) j_{\ell}[k\chi(z)] \,,
\end{equation}
where $p_z(z)$ is the normalized distribution of the sources, $b(z)$ is the linear bias, $T_{\delta}(k,z)$ is the matter over-density transfer function, and $j_{\ell}[k\chi(z)]$ is the $\ell$-th order spherical Bessel function, for a comoving distance $\chi(z)$. For our purposes, the theoretical curve will be obtained with the linear matter power spectrum for a constant linear bias $b(z)=1$. In this way, we can interpret the clustering at different scales by comparing it with the distribution of dark matter in a linear perturbation theory.

In real surveys, the sky coverage is not complete, and the observed galaxy distribution is masked by regions where observations are not possible (e.g., due to foreground contamination or survey limitations). This partial sky coverage introduces correlations between different $\ell$ modes, which must be considered in the analysis. The effect of the survey mask is described by the mixing matrix, $R_{\ell \ell'}$, which convolves the true angular power spectrum with the power spectrum of the mask 
\begin{equation}\label{eq:pseudo_cl}
    C_{\ell}^{\text{obs}} = \sum_{\ell'} R_{\ell \ell'}C_{\ell'}^{\text{true}}.
\end{equation}
Following \cite{Wu25}, we use the public code \textsc{namaster}\footnote{\url{https://github.com/LSSTDESC/NaMaster}}~\citep{Alonso19} to calculate the true angular power spectrum.

\subsection{\texorpdfstring{$\beta$}{} parameter estimation}
\label{sec:param-est}
Under the assumptions of isotropy and homogeneity, the expected behavior of the 2PACF follows a power-law distribution given by
\begin{equation}
\label{eq:power-law}
    \omega (\theta) = \left(\frac{\theta}{\theta_0} \right)^{-\beta}\,,
\end{equation}
where $\theta_0$ and $\beta$ are parameters related to the transition scale between linear and non-linear regimes, and the slope of the correlation, respectively~\citep{Peebles93, Coil13,Connolly02,Marques20,Coil13,Kurki-Suonio,Totsuji69}. 

To estimate the parameters $\theta_0$ and $\beta$, we adopted a Bayesian inference approach using Markov Chain Monte Carlo (MCMC) methods. This iterative process allows for an efficient exploration of the parameter space, leading to more reliable estimates of the posterior distributions~\citep{Trotta17}. For this analysis, we utilize the publicly available \textsc{emcee}\footnote{\url{https://emcee.readthedocs.io/en/stable/}} code~\citep{Foreman2013}.

Given a model characterized by a set of parameters $\bm{\Theta}$ and a dataset $\bm{D}$, Bayesian inference computes the posterior probability distribution, $P(\bm{\Theta}|\bm{D})$, following Bayes' theorem,
\begin{equation}
    P(\bm{\Theta}|\bm{D}) = \frac{P(\bm{D}|\bm{\Theta})P(\bm{\Theta})}{P(\bm{D})}.
\end{equation}
The logarithm of the scaled posterior distribution is expressed as
\begin{equation}
    \log P(\bm{\Theta}|\bm{D}) \propto \log P(\bm{D}|\bm{\Theta}) + \log P(\bm{\Theta}),
\end{equation}
where the likelihood can be written as 
\begin{equation}
\log P(\bm{D}|\bm{\Theta}) \propto -\frac{1}{2}\chi^{2} \,.
\end{equation}
In this work 
\begin{equation}
\begin{split}
\chi^{2} = \sum_{i,j} \left[\omega(\theta_i) - \omega^{\text{PL}}(\theta_i;\theta_0,\beta)\right] \times C_{i,j}^{-1}(\theta_i,\theta_j) \times \\
\left[\omega(\theta_j) - \omega^{\text{PL}}(\theta_j;\theta_0,\beta)\right] \,,
\end{split}    
\end{equation}
where $\omega(\theta_i)$ represents the measured correlation function, $\omega^{\text{PL}}(\theta_i;\theta_0,\beta)$ is the model prediction with power law parameters $\theta_0$ and $\beta$, and $C_{i,j}^{-1}(\theta_i,\theta_j)$ is the inverse covariance matrix of the measurements, given by equation~(\ref{eq:cov}).

For the MCMC implementation, the prior distributions for $\theta_0$ and $\beta$ were chosen based on typical values for galaxies~\citep{Franco2024, Wang13}. The priors used are detailed in Table~\ref{tab:priors}.

\begin{table}[!ht]
\centering
\caption{Prior ranges for the power-law parameters.}
\begin{tabularx}{0.8\linewidth}{>{\centering\arraybackslash}X >{\centering\arraybackslash}X}
\hline
\hline
Parameter & Prior distribution\\
\hline
$\theta_0$ & $[0.01, 0.13]$\\
$\beta$ & $[0.2, 1.0]$\\
\hline
\end{tabularx}
\label{tab:priors}
\end{table}

The parameter $\beta$, in particular, is very useful in our analyses. 
It not only defines the slope of the correlation function, but 
also serves as an estimator for quantifying the degree of matter 
clustering. 
As discussed in \cite{Franco2024}, $\beta$ provides 
a robust characterization of the clustering behavior, offering insights into the underlying large-scale structure.

\section{Analyses and results}
\label{sec:results}

In this section, we present our tomographic analysis in two redshift shells, 
Shell 1 ($0 \leq z < 0.06$) and Shell 2 ($0.06 \leq z < 0.12$), of blue galaxies. 
Specifically, our methodology to probe the matter clustering evolution is based on 
a comparison of the valuable information extracted from both shells using diverse 
statistical tools. 

In fact, a comparison of the results from one shell to the other provides information about the evolutionary process of growth of large structures that host (or are lacking of) blue galaxies. 
This because the Shell 2 maps the galaxy distribution from a younger epoch compared with the Shell 1, which maps the oldest structures in the Universe, and we intend to capture quantitatively these characteristics in our analyses. 
This study comprehends the use of diverse tools to study the matter clustering evolution. 
The statistical estimators employed in the scrutiny of the Local Universe were: 
the 2PACF (analyses done at small and large angles), the cumulative distribution function (CDF), and the angular power spectrum.

\subsection{The 2PACF: small-angle analysis}
\label{2pacf-SA}

\begin{figure*}
    \begin{minipage}{\linewidth}
        \centering
        \includegraphics[width=\linewidth]{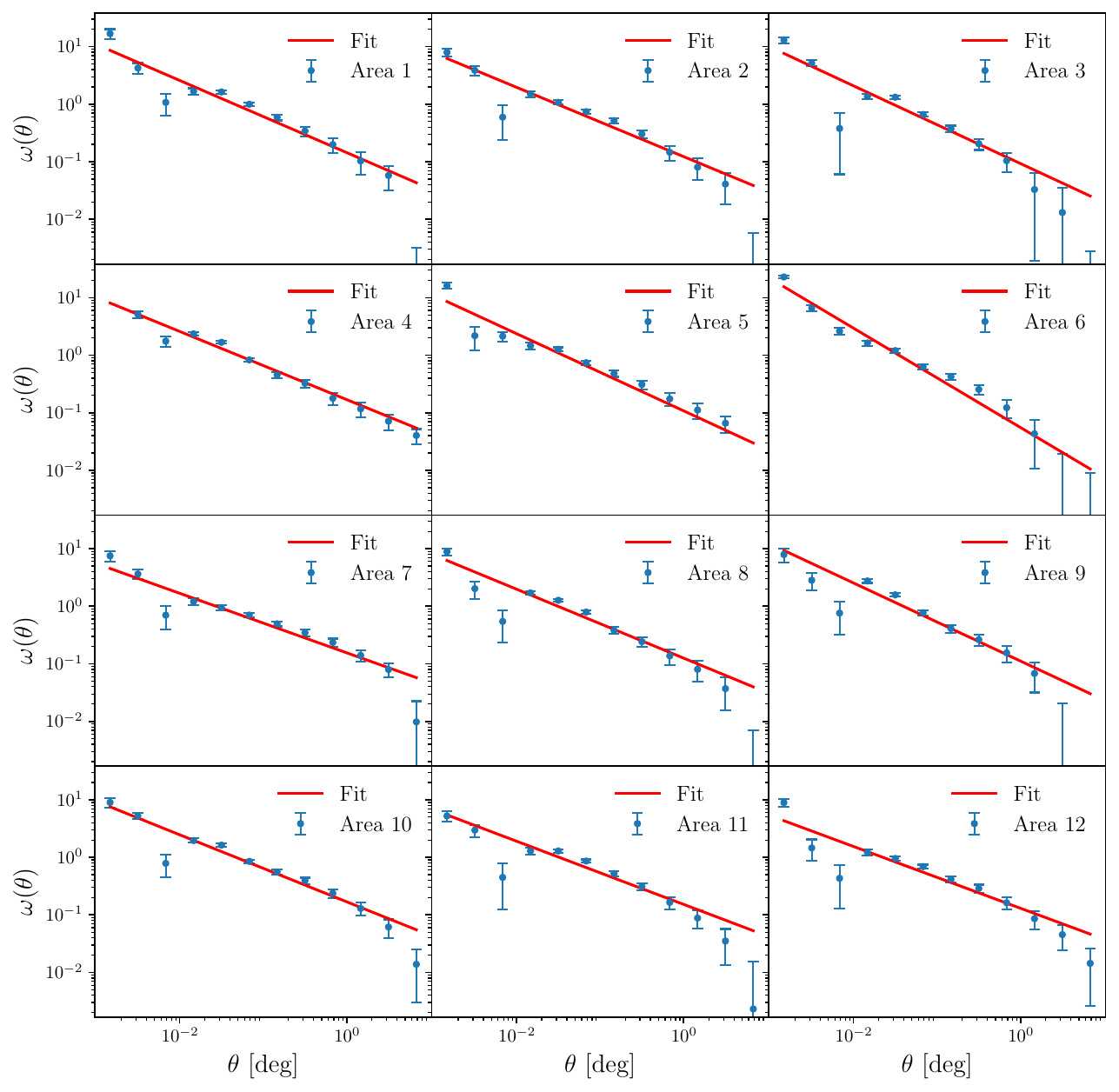}
    \end{minipage}
    \caption{2PACF for small-angle analyses in Shell 1. Angular distribution of the $12$ regions illustrated in Figure~\ref{fig:footprint_shells} within the angular range $0^{\circ} < \theta \leq 10^{\circ}$.}
    \label{fig:tpacf_shell1_small}
\end{figure*}

\begin{figure*}
    \begin{minipage}{\linewidth}
        \centering
        \includegraphics[width=\linewidth]{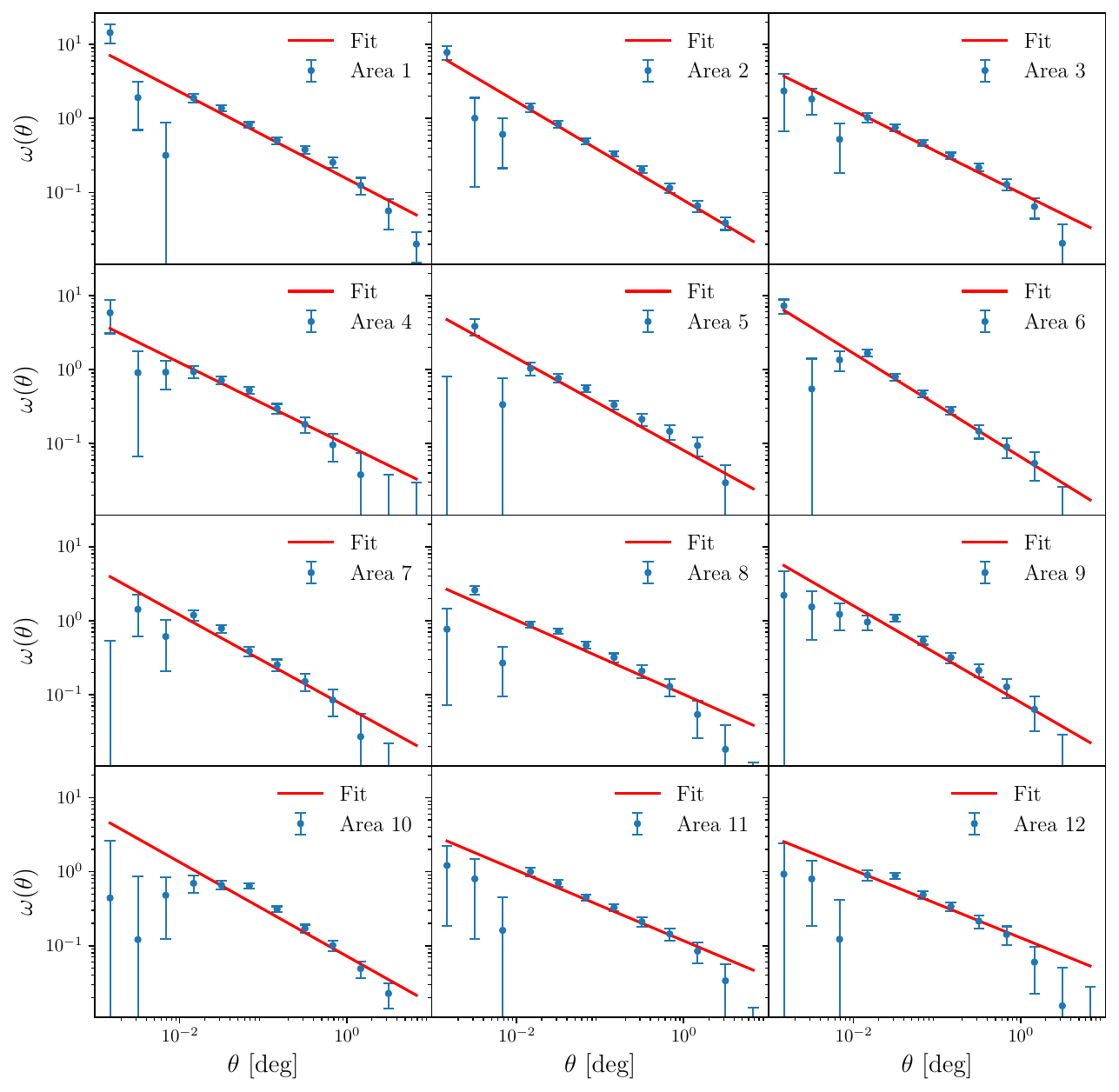}
    \end{minipage}
    \caption{2PACF for small-angle analyses in Shell 2. Angular distribution of the $12$ regions illustrated in Figure~\ref{fig:footprint_shells} within the angular range $0^{\circ} < \theta \leq 10^{\circ}$.}
    \label{fig:tpacf_shell2_small}
\end{figure*}

We start our analyses of the matter clustering studying the 2PACF, at small angles, of the data contained in the 12 Areas selected, in each one of the redshift shells (see Figure~\ref{fig:footprint_shells}). 
For each Area, we calculate the 2PACF, $\omega(\theta)$, for the angular separations $\theta \in [0^{\circ},10^{\circ}]$. 
The 2PACF binned data, with their corresponding uncertainties obtained from the mocks, are then used to adjust a best-fit power law, equation~(\ref{eq:power-law}), using the MCMC approach to find the parameters $\theta_0$ and $\beta$. 
The results for the 12 Areas, in each one of the shells, are displayed in Figures~\ref{fig:tpacf_shell1_small} and~\ref{fig:tpacf_shell2_small}, and summarized in Table~\ref{tab:best-fit_small}. 

\begin{table*}[!ht]
\centering
\caption{{\bf Small scales:} Best-fit parameters for the power-law analyses 
of the 2PACF (Figures~\ref{fig:tpacf_shell1_small} and \ref{fig:tpacf_shell2_small}): 
$\omega(\theta) = (\theta / \theta_0)^{-\beta}$. 
On average, the results for the SDSS blue galaxies are: 
$\tilde{\theta}_{0, 1} = 0.038 \pm 0.009$; $\tilde{\theta}_{0, 2} = 0.017 \pm 0.006$; 
$\tilde{\beta}_1 = 0.625 \pm 0.094$; $\tilde{\beta}_2 = 0.579 \pm 0.096$.
}
\begin{tabularx}{0.9\linewidth}{>{\centering\arraybackslash}X >{\centering\arraybackslash}X >{\centering\arraybackslash}X >{\centering\arraybackslash}X >{\centering\arraybackslash}X}
\hline
\hline
& \multicolumn{2}{c}{Shell 1} & \multicolumn{2}{c}{Shell 2} \\
\cmidrule(lr){2-3} \cmidrule(lr){4-5}
 & {$\theta_0$ [degrees]} & {$\beta$} & {$\theta_0$ [degrees]} & {$\beta$} \\
\midrule
Area 1 &  $0.048\,\pm\,0.004$ & $0.642\,\pm\,0.022$ & $0.033\,\pm\,0.004$ & $0.596\,\pm\,0.027$\\
Area 2 &  $0.038\,\pm\,0.003$ & $0.582\,\pm\,0.021$ & $0.020\,\pm\,0.002$ & $0.693\,\pm\,0.024$\\
Area 3 &  $0.034\,\pm\,0.002$ & $0.642\,\pm\,0.026$ & $0.016\,\pm\,0.002$ & $0.529\,\pm\,0.029$\\
Area 4 &  $0.054\,\pm\,0.004$ & $0.590\,\pm\,0.022$ & $0.019\,\pm\,0.002$ & $0.487\,\pm\,0.038$\\
Area 5 &  $0.033\,\pm\,0.002$ & $0.694\,\pm\,0.028$ & $0.014\,\pm\,0.002$ & $0.672\,\pm\,0.038$\\
Area 6 &  $0.031\,\pm\,0.002$ & $0.883\,\pm\,0.035$ & $0.021\,\pm\,0.002$ & $0.700\,\pm\,0.031$\\
Area 7 &  $0.023\,\pm\,0.002$ & $0.545\,\pm\,0.027$ & $0.014\,\pm\,0.002$ & $0.599\,\pm\,0.038$\\
Area 8 &  $0.039\,\pm\,0.003$ & $0.589\,\pm\,0.022$ & $0.011\,\pm\,0.001$ & $0.485\,\pm\,0.025$\\
Area 9 &  $0.039\,\pm\,0.003$ & $0.683\,\pm\,0.025$ & $0.019\,\pm\,0.002$ & $0.667\,\pm\,0.036$\\
Area 10 & $0.044\,\pm\,0.003$ & $0.594\,\pm\,0.021$ & $0.014\,\pm\,0.002$ & $0.645\,\pm\,0.024$\\
Area 11 & $0.044\,\pm\,0.003$ & $0.530\,\pm\,0.017$ & $0.012\,\pm\,0.002$ & $0.459\,\pm\,0.028$\\
Area 12 & $0.024\,\pm\,0.002$ & $0.525\,\pm\,0.023$ & $0.016\,\pm\,0.003$ & $0.416\,\pm\,0.029$\\
\bottomrule
\end{tabularx}
\label{tab:best-fit_small}
\end{table*}

The next step is a comparison of the $\beta$ values obtained by analysing a set of $12,000$ values obtained from $1,000$ Area-mocks for each of the $12$ regions in study and produced according to the fiducial cosmology (see Section~\ref{sec:mocks}).

\begin{figure}
\begin{minipage}[b]{\linewidth}
\centering
\includegraphics[width=0.9\textwidth]{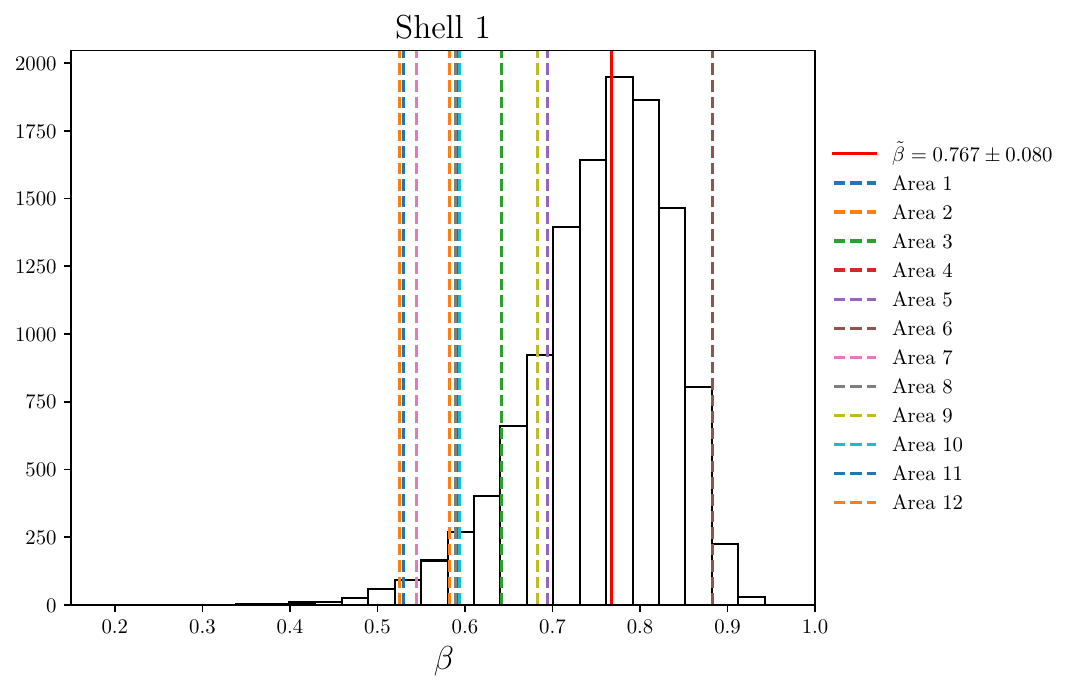}
\includegraphics[width=0.9\textwidth]{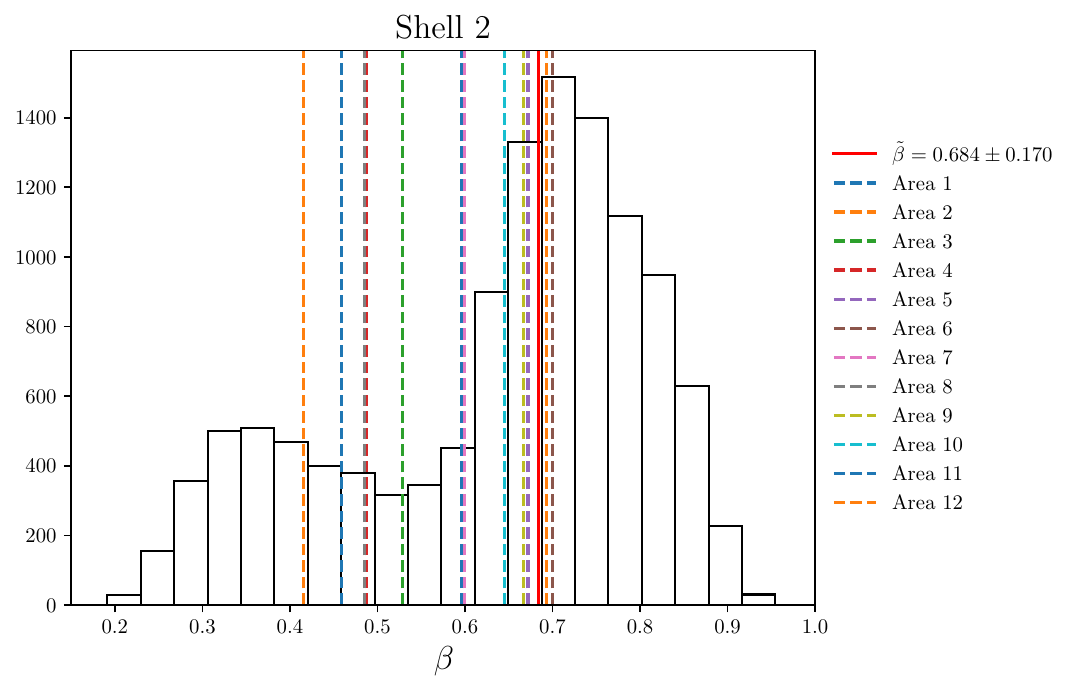}
\end{minipage}
\caption{Small-scales case.
\textbf{Upper:} Distribution of the best-fit $\beta$ parameter values from $1,000$ Area-mocks in Shell 1 ($0 \leq z < 0.06$).  
\textbf{Bottom:} Same for Shell 2 ($0.06 \leq z < 0.12$). The black histograms represent the distributions of $\beta$ values obtained from the mocks, while the vertical dashed lines indicate the $\beta$ values obtained from the data in each of the $12$ regions (cf. Table~\ref{tab:best-fit_small}), using the same analysis pipeline. The solid red line shows the median value over mocks. The median and standard deviation are $\tilde{\beta} = 0.767 \pm 0.080$ in Shell 1 and $\tilde{\beta} = 0.684 \pm 0.170$ in Shell 2.
}
\label{fig:beta-distribution_small}
\end{figure}

For each of the $12,000$ mock realizations, we fit the 2PACF, extract the best-fit $\beta$ parameter, and build a distribution of values per shell, as can be seen in Figure~\ref{fig:beta-distribution_small}. 
These distributions provide information regarding the clustering strength in each shell according to the fiducial cosmology. 
In addition, in Appendix~\ref{app:percentiles} we perform complementary analyses 
to assess a statistical comparison between the $\beta$ values from our data analyses and those values obtained from mocks. 


The behavior observed in these distributions, individually and comparatively, is consequence of the growth process of structures. 
For data from Shell 2, corresponding to a Universe younger than the structures mapped in Shell 1, one observes a bimodality\footnote{This bimodality suggests that there exist two populations of blue galaxies: one with small $\beta$ is localized in under-dense regions like voids and filaments and the other with large $\beta$ localized in over-dense regions like galaxy groups and clusters} in the $\beta$ distribution from the mocks data, suggestive of under-dense regions, as shown in~\cite{Franco2024}. 
For the Shell 1, this feature is absent, but the distribution is clearly skewed to the left, perhaps a remnant of the bimodality from the younger epoch.

From the analyses of the 12 Areas in Shell 1, and independently the 12 Areas in Shell 2, 
summarized in the distributions displayed in 
Figure~\ref{fig:beta-distribution_small}, they all show 
consistency with what is expected in the concordance 
cosmological model, represented by the outcomes from the set of mocks. 
Additionally, because larger values of $\beta$ means stronger clustered matter one expects that the data in Shell 1 be more clustered than the data in Shell 2; in fact this is corroborated in our analyses of the SDSS blue galaxies and in the data mocks: \\
$\tilde{\beta}^{\text{\sc sdss}}_1 = 0.625$ is larger than 
$\tilde{\beta}^{\text{\sc sdss}}_2 =  0.579$; and \\
$\tilde{\beta}^{\text{mocks}}_1 = 0.767$ is larger than 
$\tilde{\beta}^{\text{mocks}}_2 = 0.684$.
\subsection{The 2PACF: large-angle analysis}
\label{2pacf-LA}

\begin{figure*}
    \begin{minipage}{\linewidth}
        \centering
        \includegraphics[width=\linewidth]{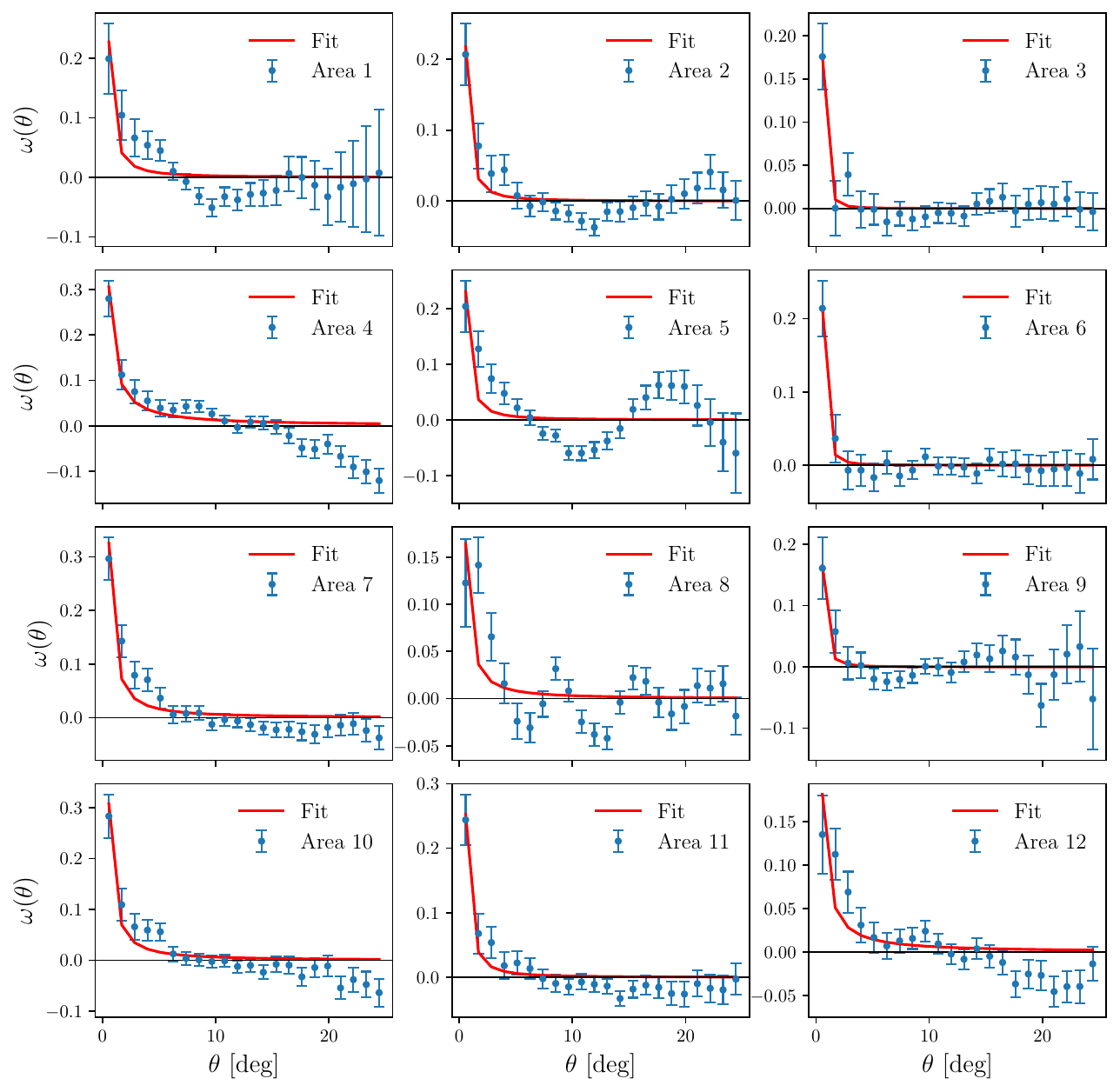}
    \end{minipage}
    \caption{2PACF for large-angle analyses in Shell 1. Angular distribution of the $12$ regions illustrated in Figure~\ref{fig:footprint_shells} within the angular range $0^{\circ} < \theta \leq 25^{\circ}$.}
    \label{fig:tpacf_shell1_large}
\end{figure*}
\begin{figure*}
    \begin{minipage}{\linewidth}
        \centering
        \includegraphics[width=\linewidth]{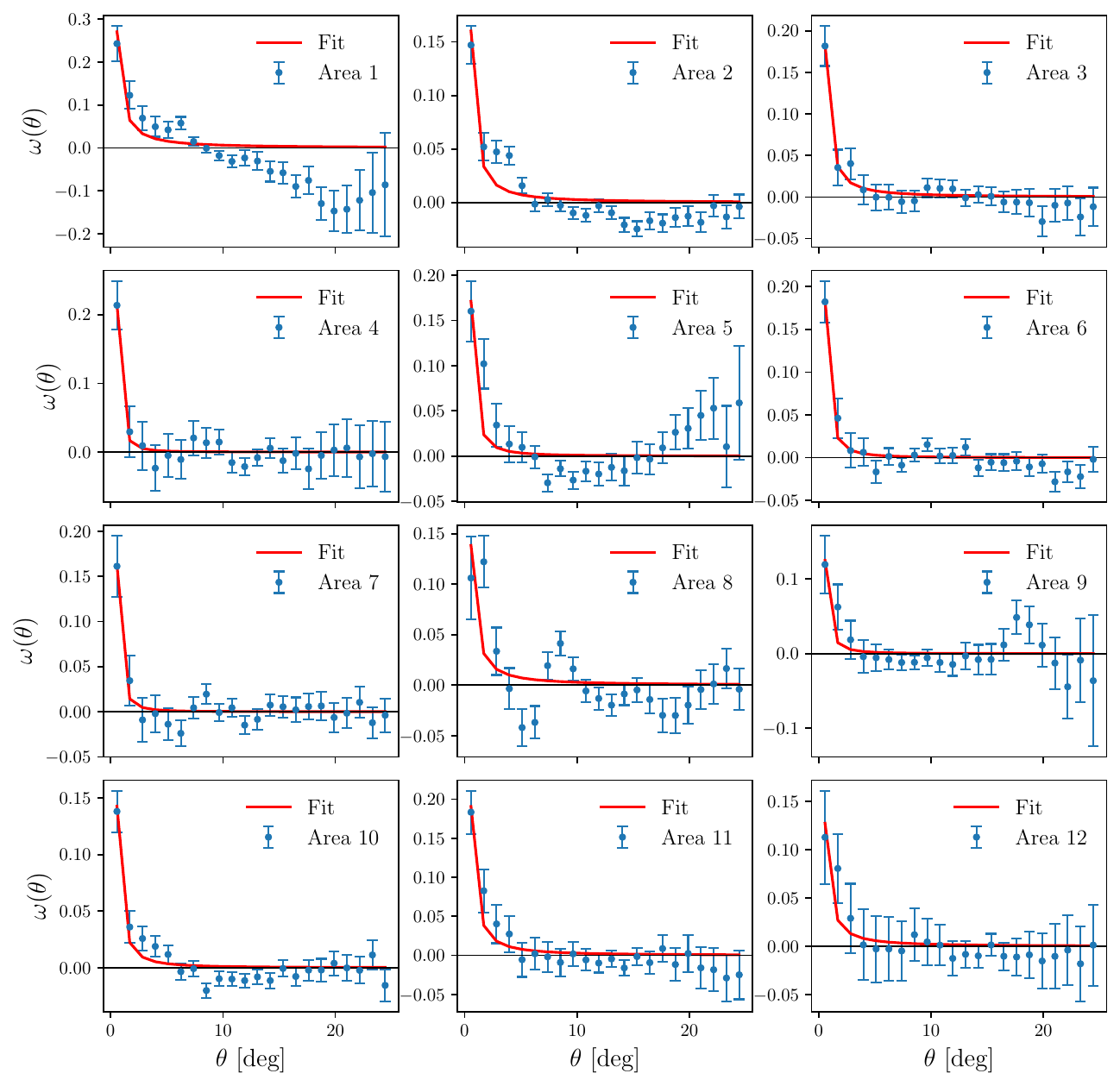}
    \end{minipage}
    \caption{2PACF for large-angle analyses in Shell 2. Angular distribution of the $12$ regions illustrated in Figure~\ref{fig:footprint_shells} within the angular range $0^{\circ} < \theta \leq 25^{\circ}$.}
    \label{fig:tpacf_shell2_large}
\end{figure*}

We now extent our 2PACF analysis to larger angular scales. This is also valuable to provide insights into the presence and features of large cosmic structures. As in the small-angle case, we analyse the 2PACF for each of the $12$ Areas in both redshift shells, this time within the angular separation range $\theta \in [0^{\circ},25^{\circ}]$. The results for the 12 Areas, in each one of the shells, are displayed in Figures~\ref{fig:tpacf_shell1_large} and~\ref{fig:tpacf_shell2_large}, and summarized in Table~\ref{tab:best-fit_large}.

\begin{table*}[!ht]
\centering
\caption{{\bf Large scales:} Best-fit parameters for the power-law analyses of the 2PACF (Figures~\ref{fig:tpacf_shell1_large} and \ref{fig:tpacf_shell2_large}): 
$\omega(\theta) = (\theta / \theta_0)^{-\beta}$. 
On average, the results for the SDSS blue galaxies are: 
$\tilde{\theta}_{0, 1} = 0.232 \pm 0.050$; $\tilde{\theta}_{0, 2} = 0.196 \pm 0.045$; 
$\tilde{\beta}_1 = 1.725 \pm 0.475$; $\tilde{\beta}_2 = 1.700 \pm 0.324$.
}
\begin{tabularx}{0.9\linewidth}{>{\centering\arraybackslash}X >{\centering\arraybackslash}X >{\centering\arraybackslash}X >{\centering\arraybackslash}X >{\centering\arraybackslash}X}
\hline
\hline
& \multicolumn{2}{c}{Shell 1} & \multicolumn{2}{c}{Shell 2} \\
\cmidrule(lr){2-3} \cmidrule(lr){4-5}
 & {$\theta_0$ [degrees]} & {$\beta$} & {$\theta_0$ [degrees]} & {$\beta$} \\
\midrule
Area 1 &  $0.22\,\pm\,0.08$ & $1.6\,\pm\,0.4$ & $0.21\,\pm\,0.04$ & $1.3\,\pm\,0.2$\\
Area 2 &  $0.24\,\pm\,0.07$ & $1.8\,\pm\,0.5$ & $0.16\,\pm\,0.03$ & $1.4\,\pm\,0.2$\\
Area 3 &  $0.29\,\pm\,0.19$ & $2.6\,\pm\,2.4$ & $0.18\,\pm\,0.05$ & $1.5\,\pm\,0.3$\\
Area 4 &  $0.19\,\pm\,0.04$ & $1.1\,\pm\,0.1$ & $0.29\,\pm\,0.15$ & $2.3\,\pm\,1.7$\\
Area 5 &  $0.24\,\pm\,0.07$ & $1.7\,\pm\,0.4$ & $0.21\,\pm\,0.08$ & $1.8\,\pm\,0.6$\\
Area 6 &  $0.31\,\pm\,0.13$ & $2.5\,\pm\,1.7$ & $0.23\,\pm\,0.07$ & $1.9\,\pm\,0.6$\\
Area 7 &  $0.25\,\pm\,0.04$ & $1.4\,\pm\,0.2$ & $0.25\,\pm\,0.13$ & $2.2\,\pm\,1.4$\\
Area 8 &  $0.15\,\pm\,0.07$ & $1.4\,\pm\,0.3$ & $0.13\,\pm\,0.07$ & $1.4\,\pm\,0.4$\\
Area 9 &  $0.26\,\pm\,0.17$ & $2.3\,\pm\,1.8$ & $0.20\,\pm\,0.14$ & $2.0\,\pm\,1.3$\\
Area 10 & $0.24\,\pm\,0.04$ & $1.4\,\pm\,0.2$ & $0.18\,\pm\,0.05$ & $1.7\,\pm\,0.3$\\
Area 11 & $0.26\,\pm\,0.06$ & $1.7\,\pm\,0.4$ & $0.18\,\pm\,0.06$ & $1.5\,\pm\,0.4$\\
Area 12 & $0.13\,\pm\,0.05$ & $1.2\,\pm\,0.2$ & $0.13\,\pm\,0.11$ & $1.4\,\pm\,0.7$\\
\bottomrule
\end{tabularx}
\label{tab:best-fit_large}
\end{table*}

\begin{figure}
\begin{minipage}[b]{\linewidth}
\centering
\includegraphics[width=0.9\textwidth]{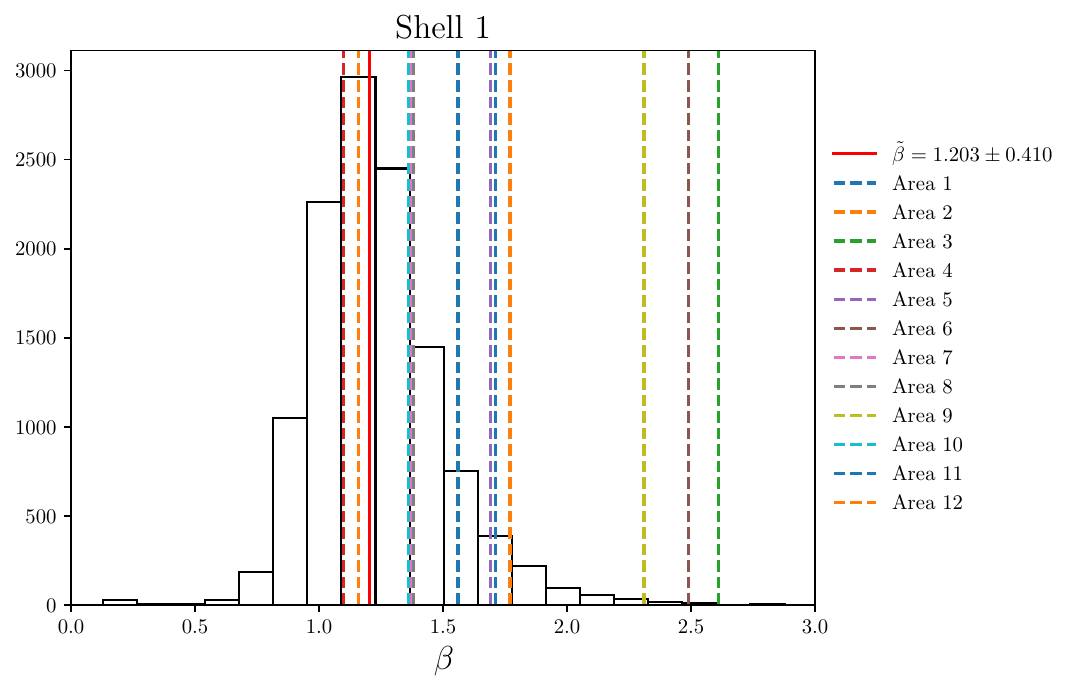}
\includegraphics[width=0.9\textwidth]{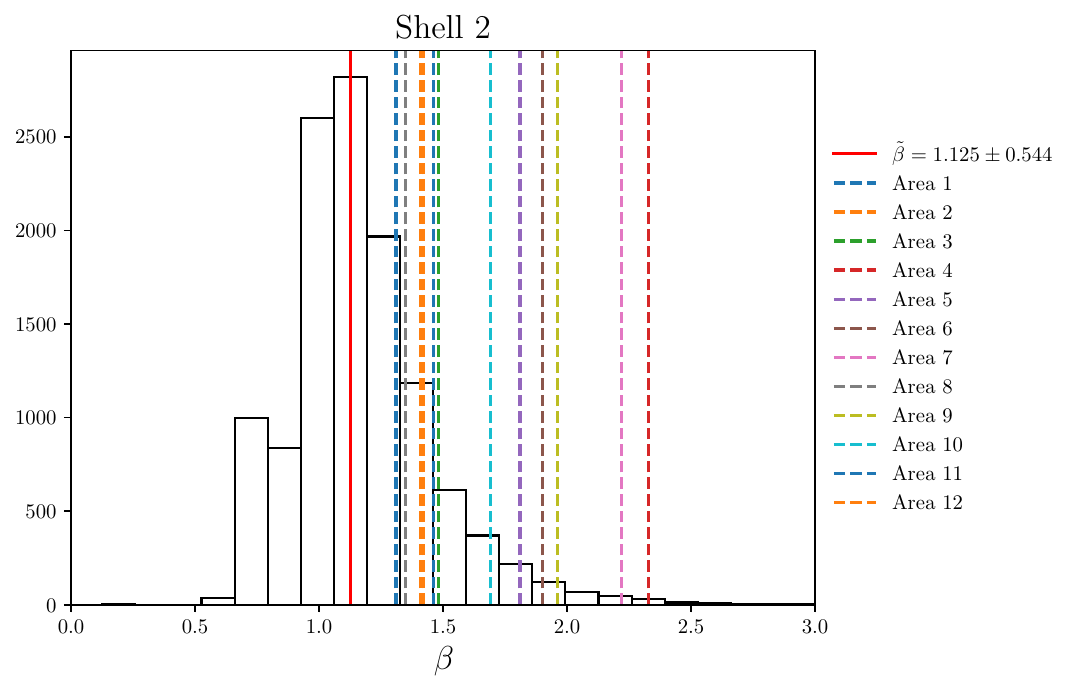}
\end{minipage}
\caption{
Histograms of the best-fit $\beta$ values, similar to Figure~\ref{fig:beta-distribution_small} but for large scales. 
The solid lines show the median values of the mocks. 
%
\textbf{Upper:} The median and standard deviation are $\tilde{\beta} = 1.203 \pm 0.410$ in Shell 1.
\textbf{Bottom:} The median and standard deviation are $\tilde{\beta} = 1.125 \pm 0.544$ in Shell~2.
}
\label{fig:beta-distribution_large}
\end{figure}

The comparison between the $\beta$ values from the SDSS blue galaxies data with the distribution obtained from the set of mock realizations are presented in Figure~\ref{fig:beta-distribution_large}, and complemented with the Table~\ref{tab:best-fit_large}. 
Moreover, we perform complementary analyses to assess a statistical comparison 
between the $\beta$ values from the our data analyses and those values obtained 
from the mocks; this study is shown in 
Appendix~\ref{app:percentiles}.

In the large-scale analyses, however, the 2PACF calculated for each Area is more informative than the support offered by the parameters $(\theta_0,\beta)$. 
This because one learns to interpret features observed in the 2PACF as signatures due to the presence of over-dense and under-dense regions in the sample in study, with the obvious caution that this information is relative to the projected 
data~\citep{Franco2024}. 

The 2PACF, displayed in Figure~\ref{fig:tpacf_shell1_large} for the Areas in Shell~1, and Figure~\ref{fig:tpacf_shell2_large} for the Areas in Shell 2, show distinctive characteristics of the blue galaxies clustering in each Area and each shell. 
Regarding the signatures shown by these 2PACF, one clearly distinguish three patterns: 
(i) a flat curve, with small fluctuations around zero (e.g., Areas 6 and 7 from Shell 1); 
(ii) fluctuation of the data points around zero but with an excess of several points over zero (e.g., Area 8 from Shell 1 and Shell 2); 
(iii) fluctuation of the data but with a defect, i.e., a valley, of several data points under zero (e.g., Area~5 in Shell 1 and Area 1 in Shell 2). 

\begin{figure*}
\begin{minipage}[b]{\linewidth}
\centering
\includegraphics[width=0.45\textwidth]{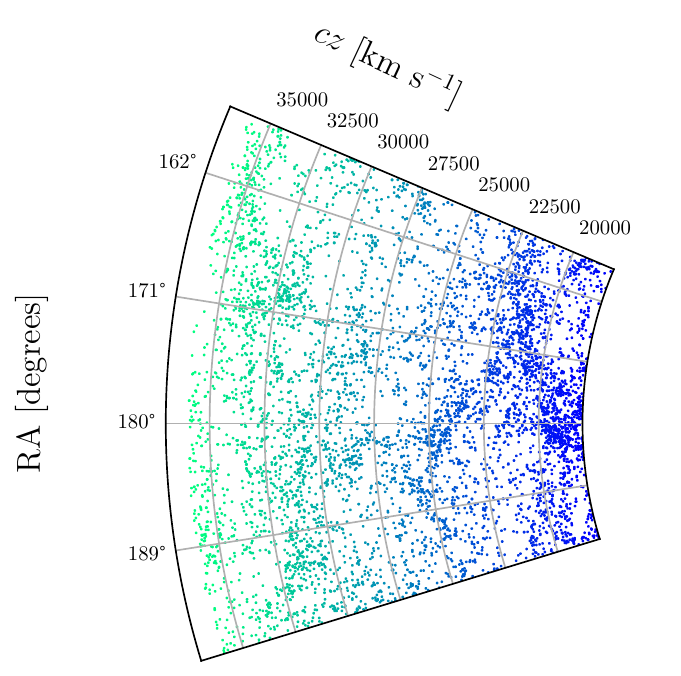}
\includegraphics[width=0.45\textwidth]{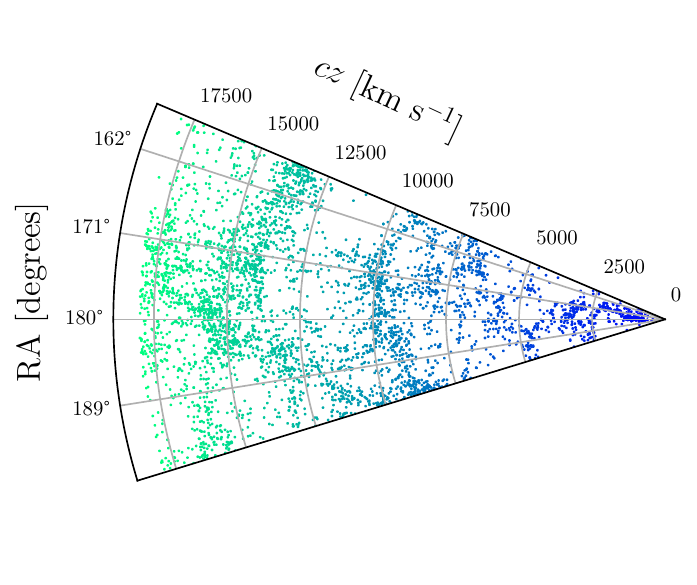}

\includegraphics[width=0.45\textwidth]{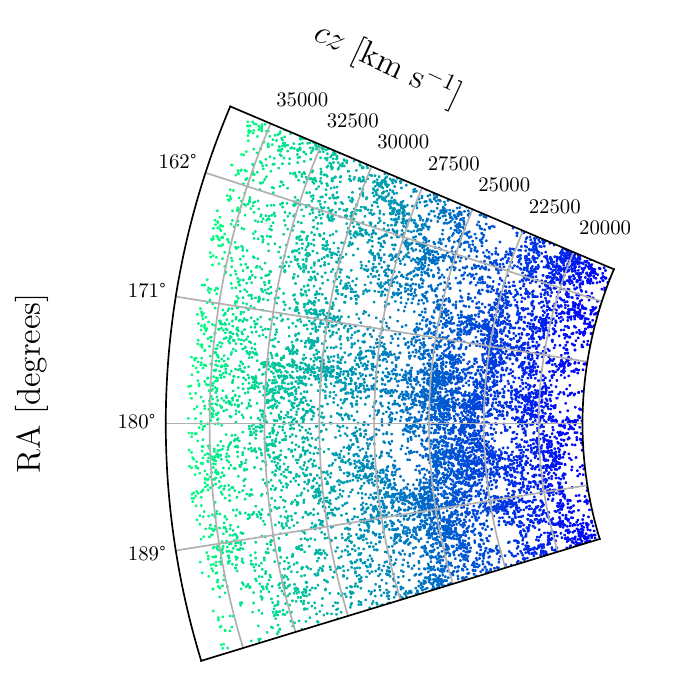}
\includegraphics[width=0.45\textwidth]{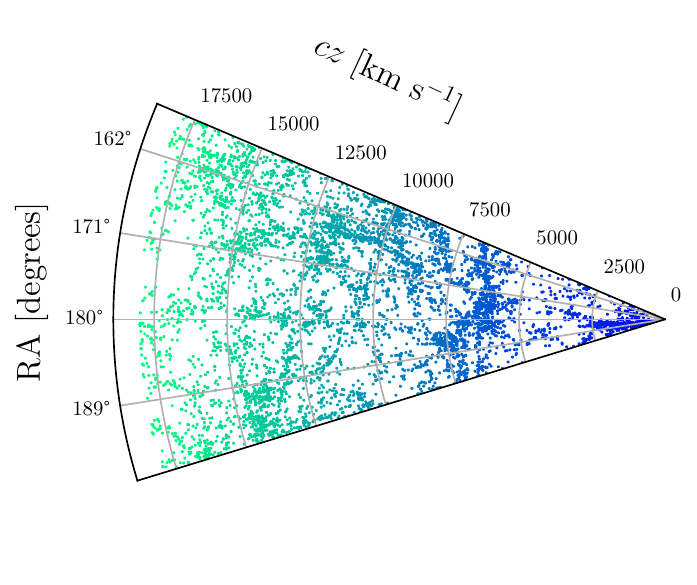}
\end{minipage}
\caption{Distribution of galaxies from Shell 2 (left) and Shell 1 (right). These plots include physical distances of the blue galaxies to the origin, in units of km s$^{-1}$. \textbf{Upper panel}: Wedge plot of Area 5. We notice the presence of bulges of blue galaxies at several distances from the origin, data that appear as clustered matter when projected on the sky (i.e., Area 5).
\textbf{Bottom panel}: Wedge plot of Area 8. We notice the presence of much more blue galaxies than in the wedge plot of Area 5 (see the Table~\ref{tab:features-areas} for the observational features of the 12 Areas in study). Again, large concentrations of galaxies are observed at several distances from the origin, data that also appear clustered when projected on the sky (i.e., Area 8).}
\label{fig:wedge_plot_a5_a8}
\end{figure*}

To better understand what the origin of these signatures could be, we use complementary information provided by wedge plots and CDF plots (for the application of the CDF in this type of analysis see, e.g.,~\cite{Franco2024}). 
As observed in Figure~\ref{fig:wedge_plot_a5_a8}, the wedge plots of Area 5 and Area 8, interesting clustered patterns appear. 
The structures observed in the wedge plot are consequence of the evolutionary process of structures growth along time. 

\begin{figure*}
    \begin{minipage}[b]{\linewidth}
        \centering
        \includegraphics[width=0.24\textwidth]{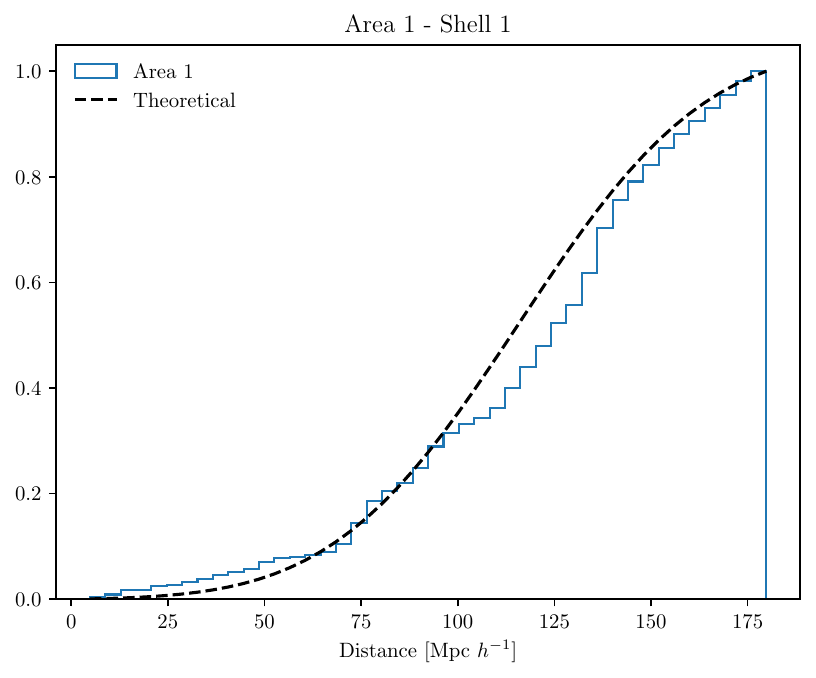}
        \includegraphics[width=0.24\textwidth]{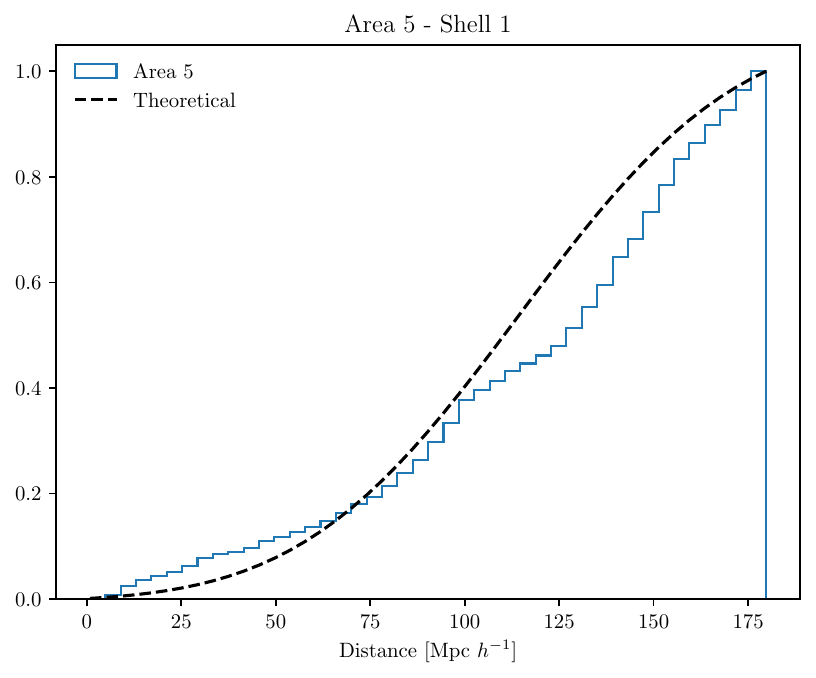}
        \includegraphics[width=0.24\textwidth]{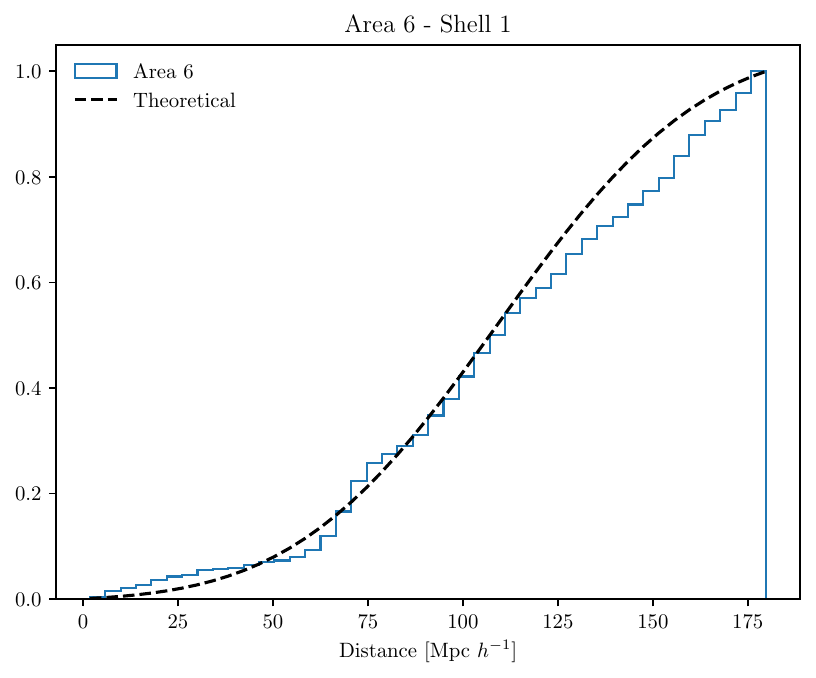}
        \includegraphics[width=0.24\textwidth]{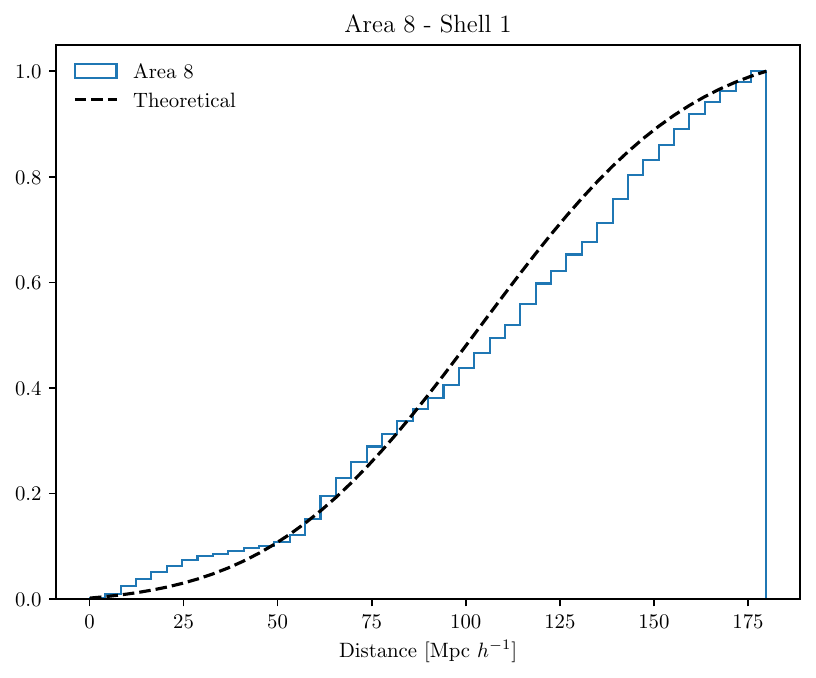}
        
        \includegraphics[width=0.24\textwidth]{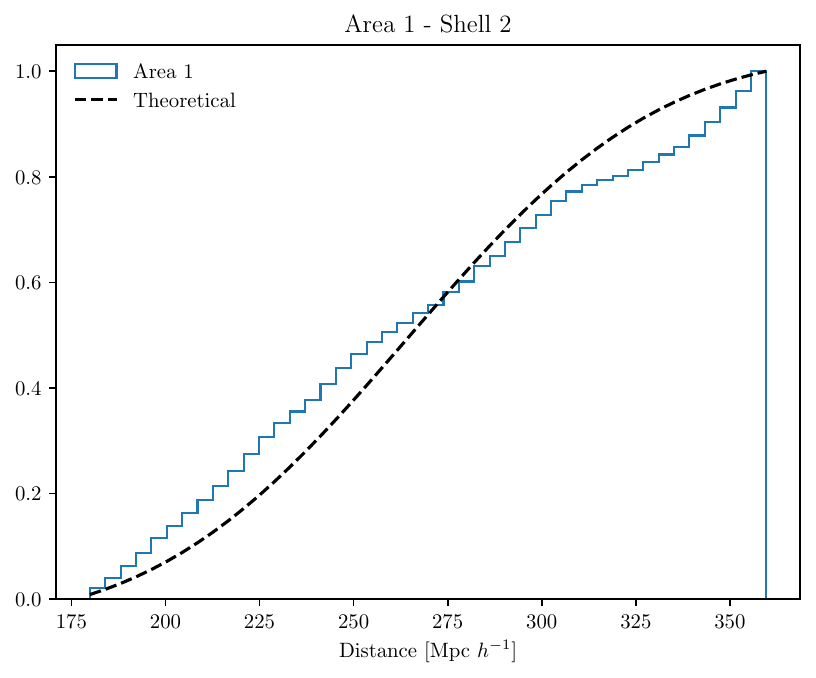}
        \includegraphics[width=0.24\textwidth]{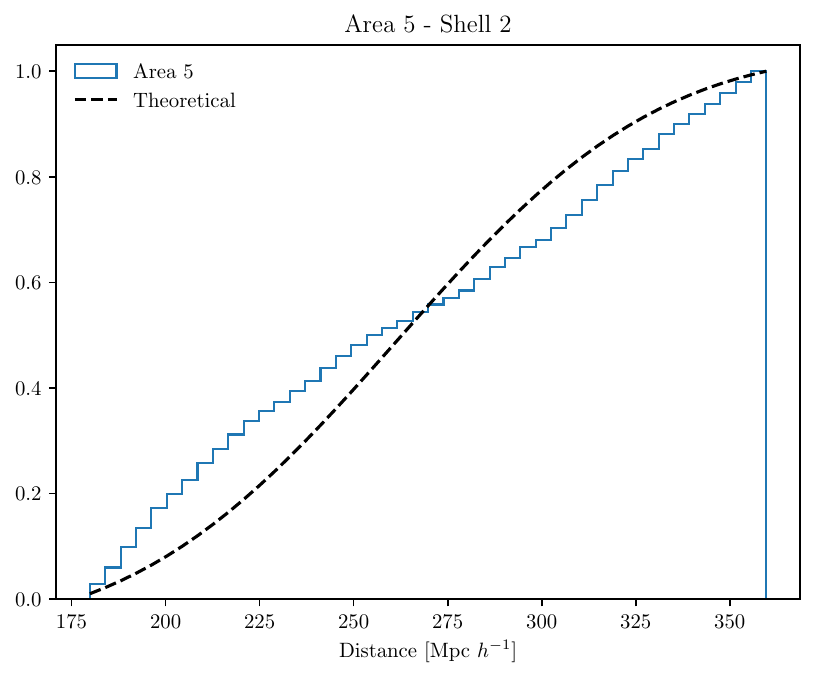}
        \includegraphics[width=0.24\textwidth]{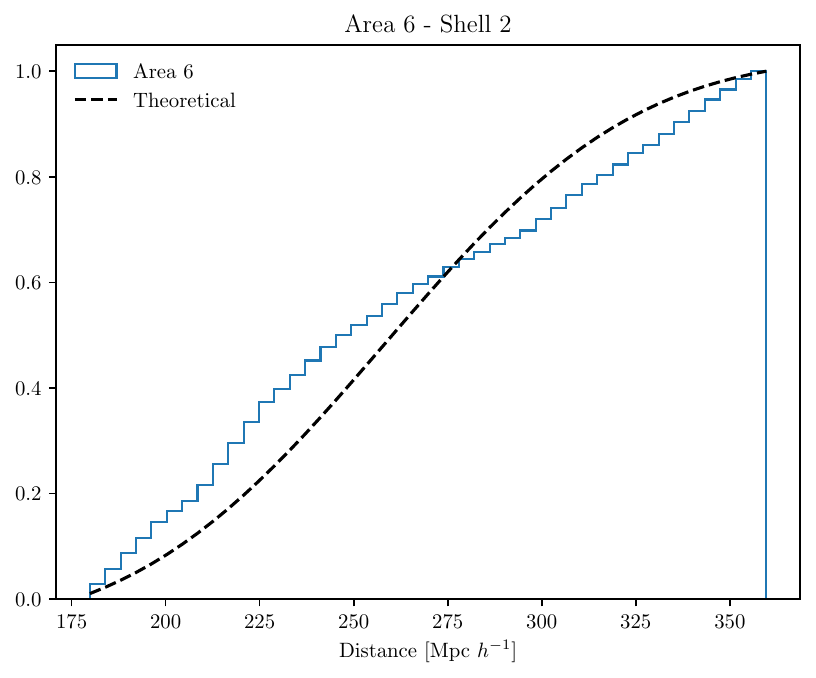}
        \includegraphics[width=0.24\textwidth]{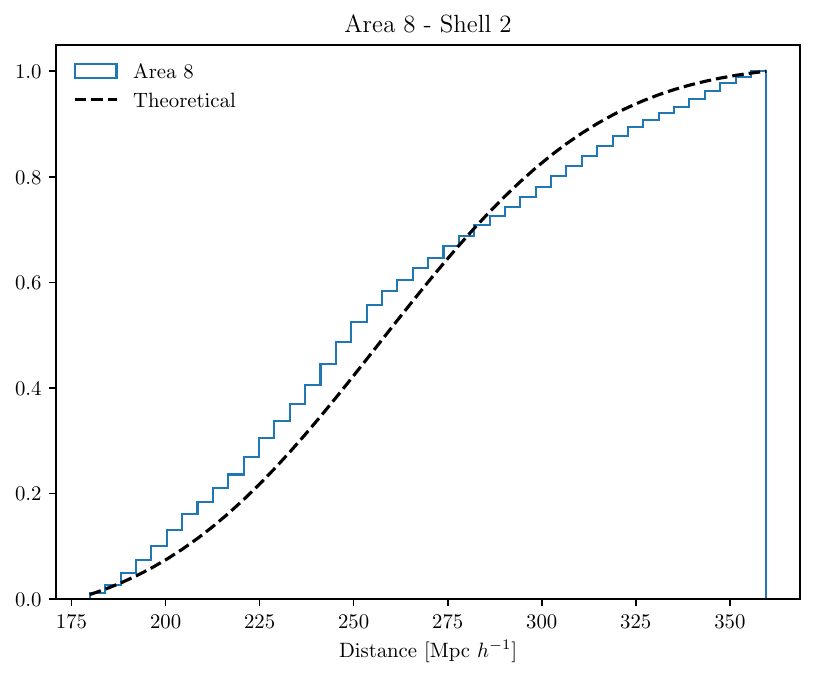}        
\end{minipage}
\caption{The first (second) row shows the CDF analysis for the Areas 1, 5, 6, 8 from the Shell 1 (Shell 2). 
Differences among different Areas in each shell are noticed. 
But, Areas in a given shell, which means structures with the same age, notably follows the same pattern, namely regions of similar scales with strong clustering followed by regions with lack of galaxies (voids). 
All these features appear independent of the number galaxies in each Area, $N_i$, as can be corroborated in the Figure~\ref{fig:wedge_plot_a5_a8}, which presents the wedge plots of Area 5 with $N_5 = 4,292$ and $5,921$, and Area 8 with $N_8 = 6,170$ and $12,144$, for Shell~1 and 
Shell~2, respectively (see Table~\ref{tab:features-areas}). 
Notice that the 2PACF of these four Areas show different signatures (see Figures~\ref{fig:tpacf_shell1_large} and~\ref{fig:tpacf_shell2_large}).
}
\label{fig:cdf}
\end{figure*}

On the other hand, the CDF plots from Areas 1, 5, 6, 8, both shells, shown in Figure~\ref{fig:cdf} exhibit other type of information. 
One can observe differences among these CDF plots for different Areas in each shell. 
But, Areas in a given shell, which means structures with the same age, notably follows the same pattern, namely regions of similar scales with strong clustering followed by regions with lack of galaxies (voids). 
The presence of these features appears independent of the number galaxies in each Area, $N_i$, as can be corroborated in the Figure~\ref{fig:wedge_plot_a5_a8}, which presents the wedge plots of Area 5 with $N_5 = 4,292$ and $5,921$, and Area 8 with $N_8 = 6,170$ and $12,144$, for Shell~1 and 
Shell~2, respectively (see Table~\ref{tab:features-areas}). 

As in small scales, on large scales one can also use the $\beta$ parameter to compare the clustering strength in different Universe epochs, namely 
in Shell~1 versus in Shell~2. 
In fact, as expected, Shell 1 exhibits stronger clustering than Shell 2, again confirmed both in the SDSS data in study and in the mock catalogs: \\
$\tilde{\beta}^{\text{\sc sdss}}_1 = 1.725$ is larger than 
$\tilde{\beta}^{\text{\sc sdss}}_2 =  1.700$; and \\
$\tilde{\beta}^{\text{mocks}}_1 = 1.203$ is larger than 
$\tilde{\beta}^{\text{mocks}}_2 = 1.125$.

To quantify the comparison between the observational and simulated $\beta$ values, we computed, for each sky region, the median and standard deviation from the corresponding set of $1,000$ mock realizations. 
This yields the distributions of $\beta$ values for each region, along with the corresponding $1\sigma$ region, as listed in Table~\ref{tab:percentiles} (see Appendix~\ref{app:percentiles} for more details).

\begin{table*}[!ht]
\centering
\caption{Median and standard values of the $\beta$ parameter obtained from $1,000$ log-normal mock realizations in each of the $12$ regions, for both redshift shells. These values serve as a comparison between SDSS blues galaxies clustering and the expectations from the concordance cosmology. The histograms corresponding to this analysis 
are displayed in Figures~\ref{fig:hist_shell1_small} and~\ref{fig:hist_shell2_small} for the small scales, and 
Figures~\ref{fig:hist_shell1_large} and~\ref{fig:hist_shell2_large} 
for the large scales.}
\begin{tabularx}{0.9\linewidth}{>{\centering\arraybackslash}X>{\centering\arraybackslash}X >{\centering\arraybackslash}X >{\centering\arraybackslash}X>{\centering\arraybackslash}X}
\hline
\hline
& \multicolumn{2}{c}{Small scales} & \multicolumn{2}{c}{Large scales}\\
\cmidrule(lr){2-3} \cmidrule(lr){4-5} 
& Shell 1 & Shell 2 & Shell 1 & Shell 2\\
\midrule
Area 1 & $0.795 \pm 0.070$ & $0.702 \pm 0.065$ & $1.145 \pm 0.238$ & $1.149 \pm 0.181$\\
Area 2 & $0.762 \pm 0.076$ & $0.823 \pm 0.063$ & $1.200 \pm 0.158$ & $1.463 \pm 0.307$\\
Area 3 & $0.759 \pm 0.082$ & $0.575 \pm 0.094$ & $1.284 \pm 0.241$ & $1.071 \pm 0.180$\\
Area 4 & $0.748 \pm 0.083$ & $0.382 \pm 0.071$ & $1.501 \pm 0.292$ & $1.047 \pm 0.078$\\
Area 5 & $0.788 \pm 0.071$ & $0.707 \pm 0.061$ & $1.153 \pm 0.150$ & $1.148 \pm 0.112$\\
Area 6 & $0.761 \pm 0.078$ & $0.703 \pm 0.055$ & $1.457 \pm 0.263$ & $1.384 \pm 0.223$\\
Area 7 & $0.755 \pm 0.083$ & $0.707 \pm 0.069$ & $1.289 \pm 0.171$ & $1.229 \pm 0.140$\\
Area 8 & $0.756 \pm 0.079$ & $0.760 \pm 0.075$ & $0.931 \pm 0.076$ & $1.006 \pm 0.077$\\
Area 9 & $0.792 \pm 0.070$ & $0.700 \pm 0.060$ & $1.159 \pm 0.193$ & $1.132 \pm 0.138$\\
Area 10 & $0.770 \pm 0.079$ & $0.815 \pm 0.068$ & $1.221 \pm 0.169$ & $1.453 \pm 0.308$\\
Area 11 & $0.756 \pm 0.083$ & $0.454 \pm 0.085$ & $1.245 \pm 0.193$ & $0.929 \pm 0.085$\\
Area 12 & $0.755 \pm 0.079$ & $0.315 \pm 0.060$ & $0.988 \pm 0.093$ & $0.716 \pm 0.033$\\
\bottomrule
\end{tabularx}
\label{tab:percentiles}
\end{table*}

In fact, the values in Table~\ref{tab:percentiles} encapsulate the statistical variance expected in each region considering the mock distributions and reinforce the evolutionary trend discussed above. Despite the spread observed among regions, Shell 1 systematically exhibits higher $\beta$ values than Shell 2, both at small and large scales. 
This behavior is consistent with the 
prediction --in the concordance cosmology-- that structures in the nearby Universe are strongly clustered, 
whereas the slightly younger epoch mapped by Shell 2 displays weaker clustering. The agreement of these results with the mock distributions, shown in Appendix~\ref{app:percentiles}, and their similarity to the global patterns in Figures~\ref{fig:beta-distribution_small}~and~\ref{fig:beta-distribution_large}, support the interpretation that these differences arise from genuine structure growth throughout the evolution of the Universe 
rather than statistical fluctuations.

Additionally, these variations, although within what is expected according to the concordance cosmological model, emphasize that the clustering evolution is not equal across the sky but rather reflects the interplay between over-dense and under-dense structures.

\subsubsection{Revealing the signature of galaxy groups and clusters}
\label{sec:robustness}

One efficient way to confirm if galaxies clustered in small or large groups (i.e., galaxy clusters) are leaving a signature in the 2PACF, is by performing small angular shifts in their positions and then redoing the 2PACF to observe whether the signature decreases, disappears or nothing happens. 
In fact, one observes in the 2PACF of Area 8 a set of lumps and valleys at diverse angular scales, 
suggestive of (large) groups of galaxies~\citep{Einasto2001}, that is, the 2PACF is revealing the presence of various galaxy groups or clusters in that region. 

\begin{figure*}
\begin{minipage}[b]{\linewidth}
\centering
\includegraphics[width=0.3\textwidth]{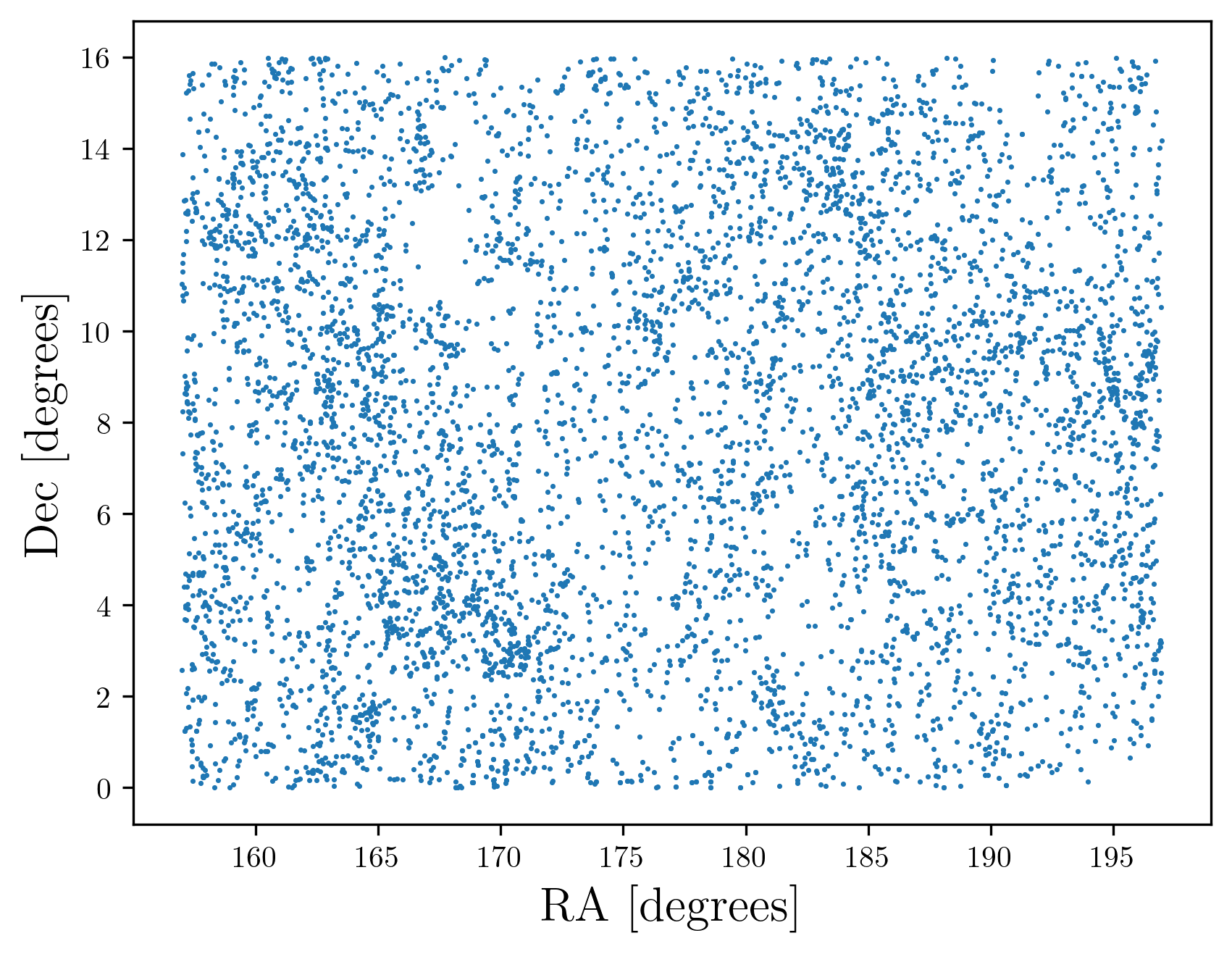}
\includegraphics[width=0.3\textwidth]{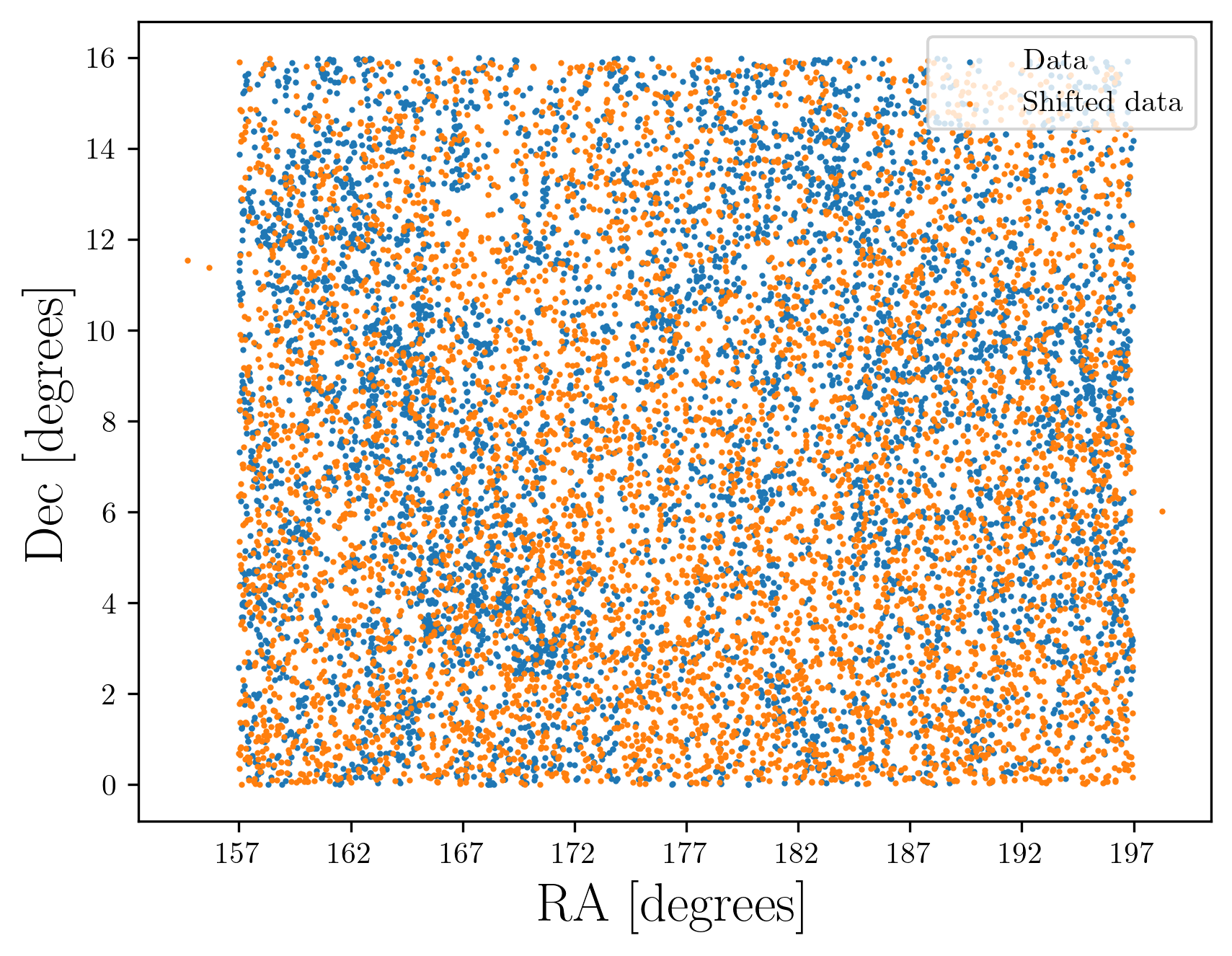}
\includegraphics[width=0.3\textwidth]{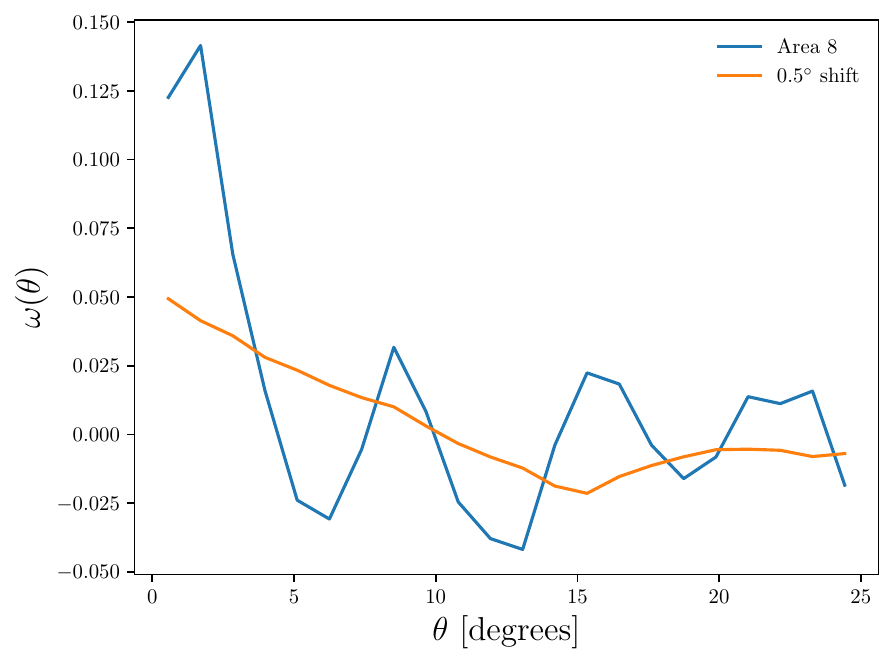} \\
\includegraphics[width=0.3\textwidth]{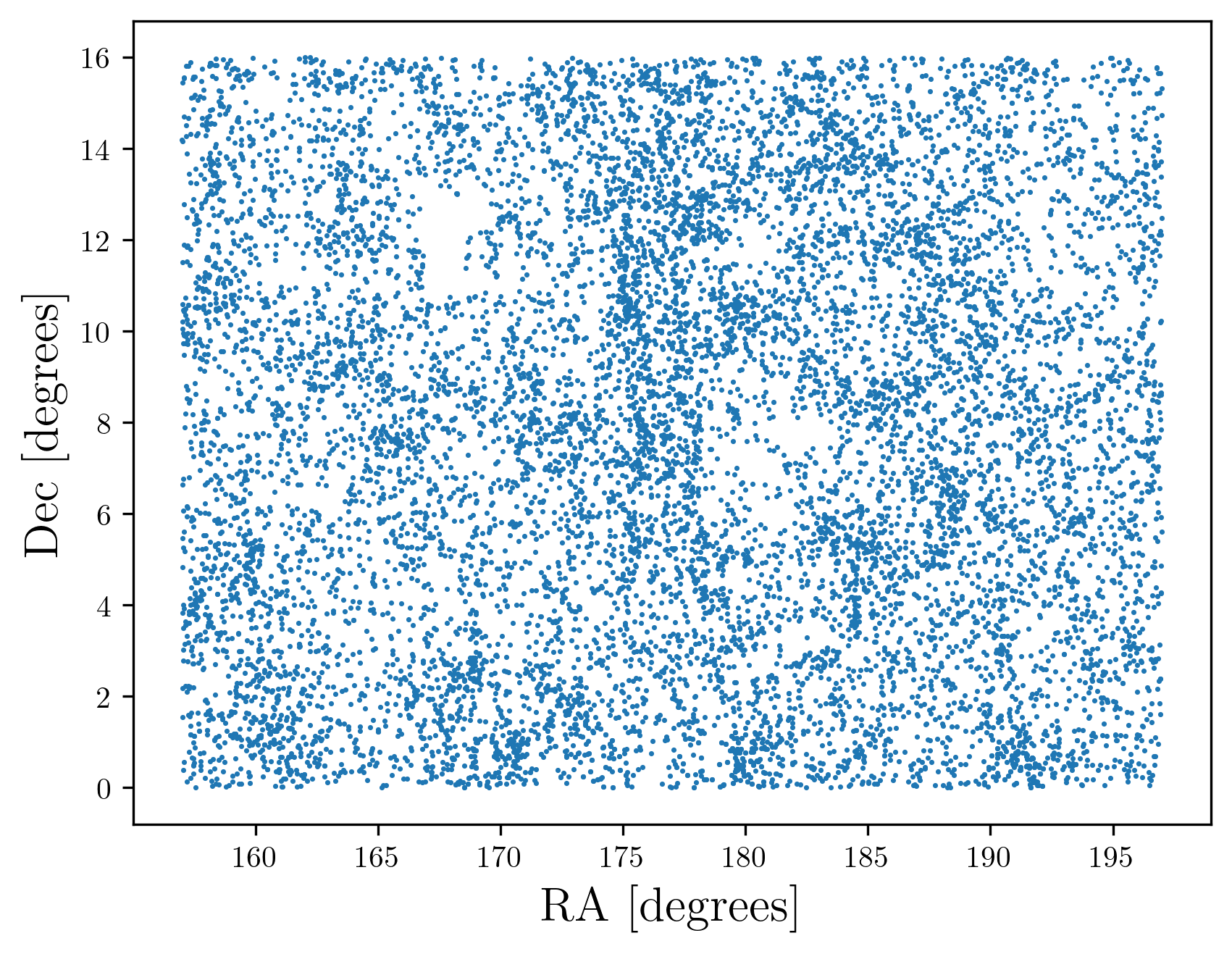}
\includegraphics[width=0.3\textwidth]{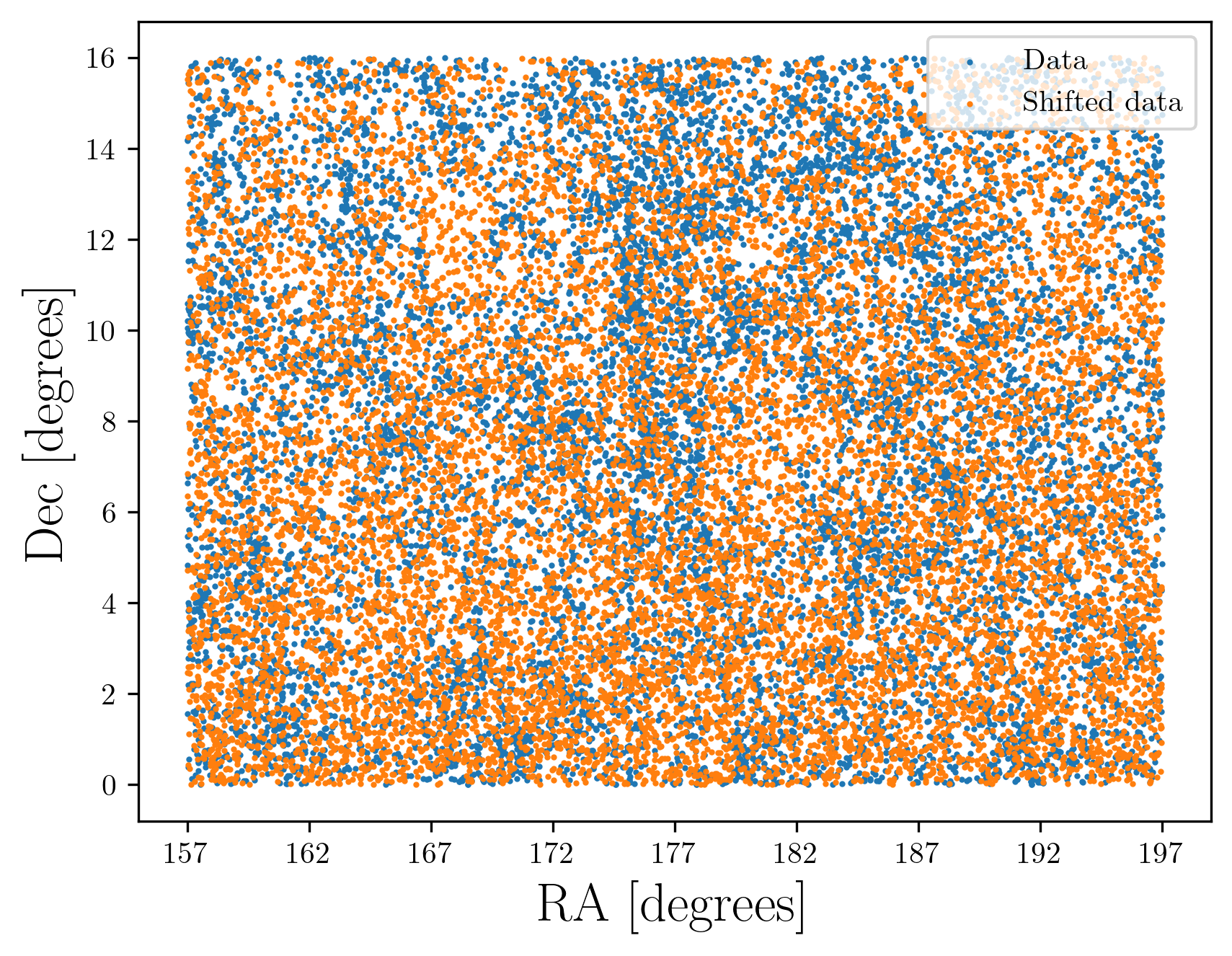}
\includegraphics[width=0.3\textwidth]{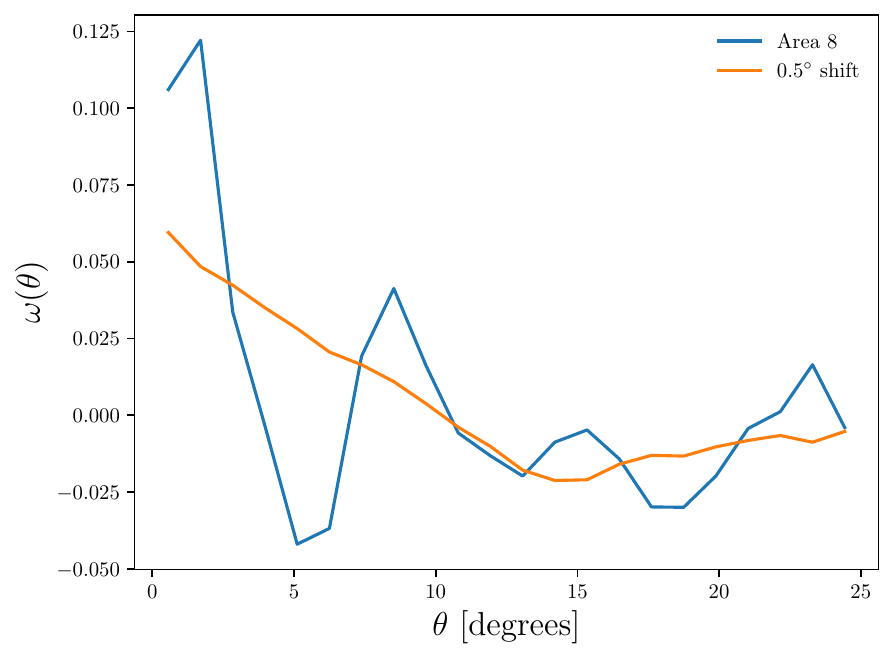}
\end{minipage}
\caption{Consistency analysis for the clustered structures mapped in Area 8. In both shells, the 2PACF from Area 8 reveals a set of lumps and valleys at diverse angular scales. 
If they correspond to galaxies concentrated in small groups or galaxy clusters then the shuffling of their angular positions will tend to destroy these features. 
\textbf{First row}: We display the Area 8 from Shell 1. 
The sequence of pictures, from left to right, is: (i) the original 2D blue galaxies distribution; (ii) these blue galaxies overlapped with their shuffled distribution (see the text for details); (iii) the 2PACF study for both distributions, i.e., analyses of the original distribution together with that one from the shuffled distribution. 
\textbf{Second row}: We display the Area 8 from Shell 2. The sequence of pictures follows the same procedures explained for the panels in the first row.} 
\label{fig:shift_a8}
\end{figure*}

A shuffling procedure is then applied to Area 8, 
where the new angular coordinates RA and Dec of each blue galaxy are randomly selected from a Gaussian distribution where the mean value is the original coordinate and the standard deviation is $0.5^{\circ}$ 
(this scale corresponds to the double 
value of $\theta_0$ obtained in Table~\ref{tab:best-fit_large}, where for lower scales non-linearities dominate). 
The panels shown in Figure~\ref{fig:shift_a8} presents the result obtained applying this shuffling procedure. 
If they correspond, indeed, to galaxies concentrated in small groups or galaxy clusters then the shuffling of their angular positions will tend to destroy these features. 
In Figure~\ref{fig:shift_a8}, we display the results of this shuffling procedure done in both shells of Area 8. 
The sequence of panels, from left to right, is: 
(i) the original 2D blue galaxies distribution; 
(ii) these blue galaxies overlapped with their shuffled distribution; 
(iii) the 2PACF study for both distributions, i.e., analyses of the original distribution together with that one from the shuffled distribution. 

Our conclusion is that, in fact, Area 8 contains large groups of galaxies. 
But perhaps more importantly, the analysis of the 2PACF of Area 8 clearly illustrates the expected difference between two distinct epochs: 
the Shell 2 (snapshot of a less evolved Universe) shows 
less galaxy groups than the Shell 1 (snapshot of a more clustered Universe).

\subsection{Angular Power Spectrum analysis}
\label{sec:Cl_results}

To compute the angular power spectrum for the selected shells with the SDSS blue galaxies, we use the~\textsc{healpix}\footnote{\url{https://healpix.sourceforge.io/}} pixelation scheme~\citep{Gorski05}, provided by the \textsc{healpy}\footnote{\url{https://healpy.readthedocs.io/en/latest/}} library~\citep{Zonca19}. 
This tool allows us to construct a fluctuation map and 
the application of the corresponding mask. 
Using the fluctuation map and the mask, we calculate the 
$C_{\ell}^{\text{true}}$ and estimate the shot 
noise, $1/\bar{n}$ (where $\bar{n}$ is the mean galaxy number density), which is then subtracted to isolate the cosmological signal. 

\begin{figure}
\centering
\includegraphics[scale=0.41]{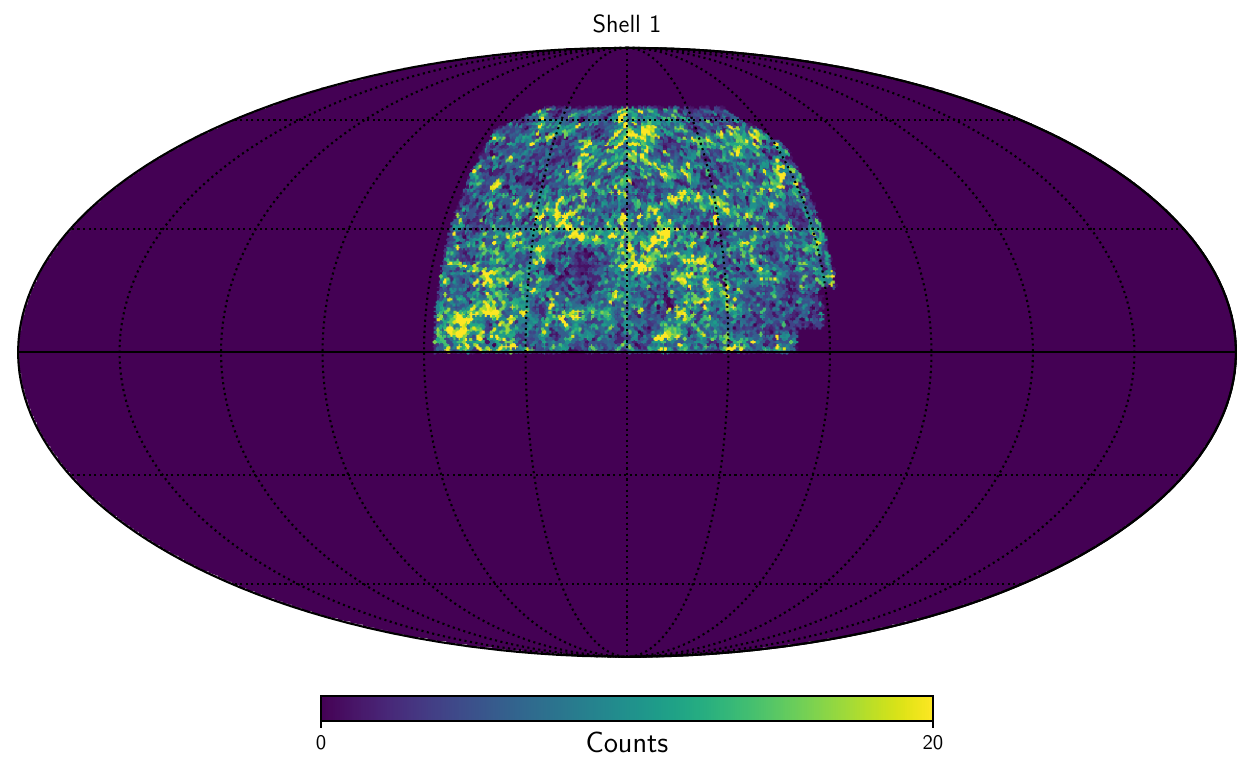}
\includegraphics[scale=0.41]{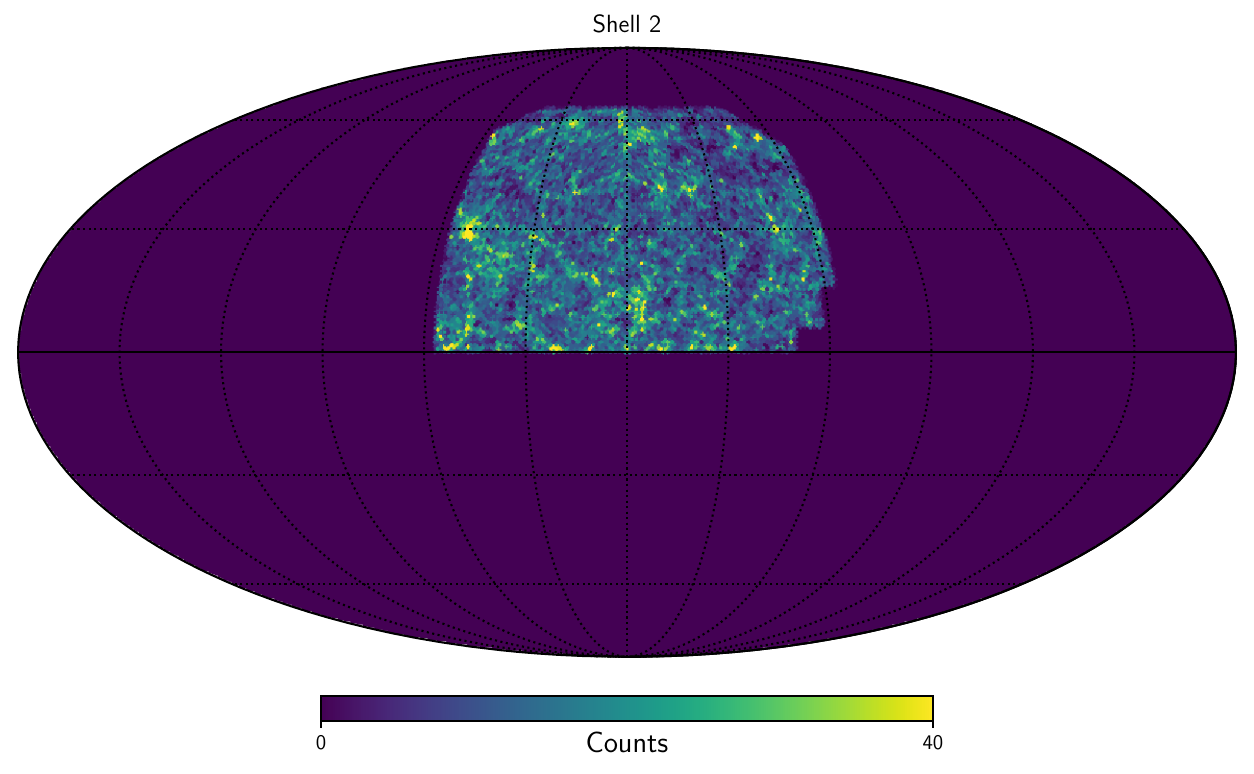}
\caption{
Number-count maps of the SDSS blue galaxies, for Shell 1 (upper map) and Shell 2 (bottom map), respectively. 
The maps were constructed using $N_{\text{side}}=64$. 
A number-count map represents the pixelized distribution of galaxies in the footprint of the data in study, 
to facilitate the computation of its angular power spectrum. 
Notice that the older part of the Universe, displayed in Shell 1, $0 \leq z < 0.06$, exhibits a more developed 
network of cosmic structures: 
clustered matter, voids, and filaments, in comparison to the scenario observed in Shell 2, $0.06 \leq z < 0.12$, where matter appears more uniformly distributed.}
\label{fig:maps_shells_64}
\end{figure}

The pixelization scheme requires a choice of pixel sizes related to the $N_{\text{side}}$ parameter, which determines the number of pixels on the sphere, given by the relationship 
\begin{equation}
N_{\text{pix}} = 12 N_{\text{side}}^{2} \,.
\end{equation}
The choice of the pixelization parameter, $N_{\text{side}}$, is crucial for our study of the matter clustering, i.e., over-dense and under-dense regions, through the analysis of the angular power spectrum. 
A low $N_{\text{side}}$ value may fail to capture the features of the galaxy distribution accurately, while a large $N_{\text{side}}$ gives rise to pixels with few galaxies, introducing bias into the estimator. 
After some tests, we choose $N_{\text{side}} = 64$ for our analyses. 
In Figure~\ref{fig:maps_shells_64}, we present the number count maps for the two selected shells. 
For comparison purposes, the analyses performed with $N_{\text{side}}=32$ are provided in the Appendix~\ref{app:healpy_32}. 
In summary, reducing $N_{\text{side}}$ from 64 to 32 implies less accuracy at small-scales, loosing information on high-$\ell$ multipoles. 
However, when comparing the two spectra, no significant differences are observed within the displayed $2 \sigma$ confidence interval.

\begin{figure}
\begin{minipage}[b]{\linewidth}
\centering
\includegraphics[width=0.9\textwidth]{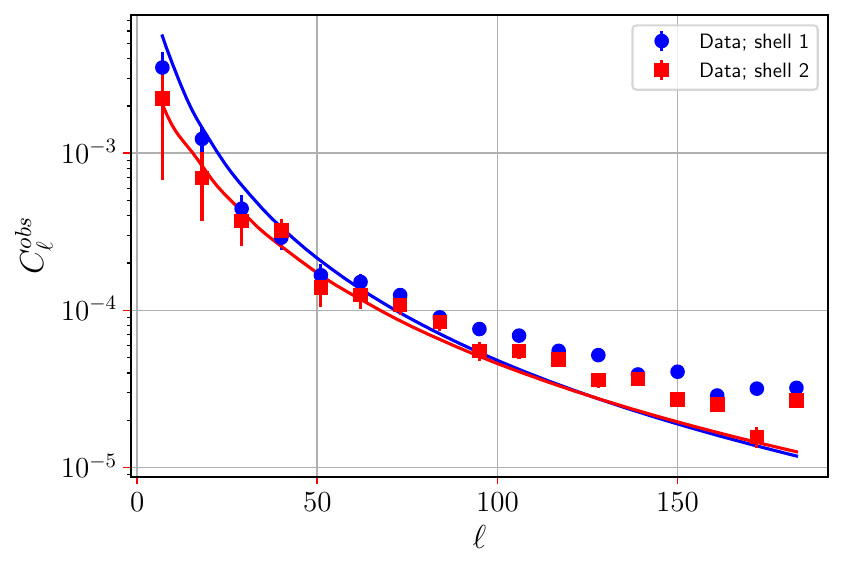}
\end{minipage}    
\caption{
True angular power spectrum, after shot-noise subtraction, for the 
data analysed in both shells. 
The multipole range is $\ell \in [7,183]$ with a bin size of $\Delta \ell = 11$. 
In this analysis, data from both shells show more power at small $\ell$, which decreases at large $\ell$. 
Additionally, at small scales $\ell \gtrsim 90$, data from Shell 1 display slightly more power than data from Shell 2, while at large scales, the power spectra in both shells are comparable.}
\label{fig:cl_shells_64}
\end{figure}

The true angular power spectrum, after shot-noise subtraction, for the two shells is presented in Figure~\ref{fig:cl_shells_64}. 
The multipole range is $\ell \in [7,183]$, with a bin size of 
$\Delta \ell = 11$. 
The vertical axis is dimensionless, as the maps represent fluctuations. 
The uncertainties were derived from the standard deviation of the results obtained from $1,000$ mock simulations, using the same methodology applied to the data. 
From a qualitative perspective, the result aligns with expectations: there is more power at small $\ell$, with a decrease in power at large $\ell$. 
Additionally, due to the evolution of the Universe, Shell 1 exhibits more power than Shell 2. 
We also notice that, because the redshift shells are 
contiguous, at large scales the power spectra are comparable. 

\begin{figure}
\begin{minipage}[b]{\linewidth}
\centering
\includegraphics[width=0.9\linewidth]{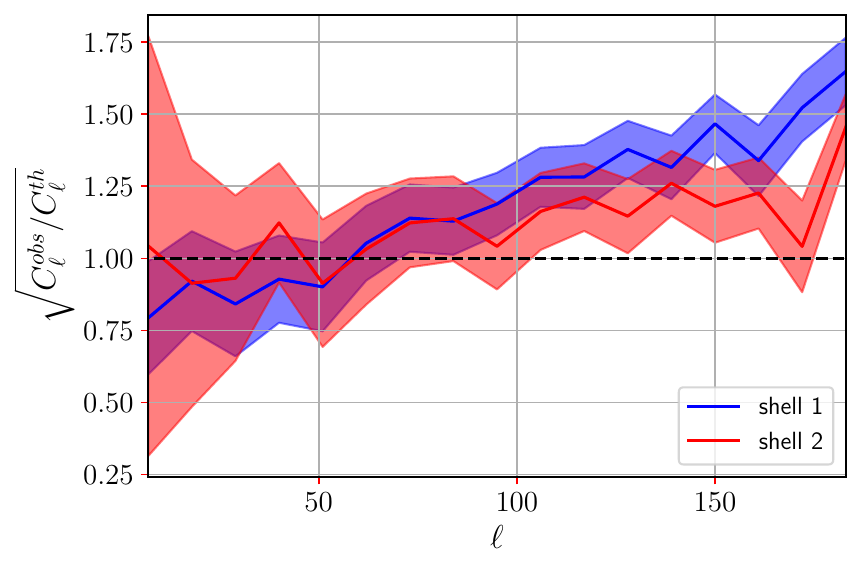}
\end{minipage}
\caption{
Comparison of the observed and theoretical angular power spectra for both shells. 
The shaded region represents $2 \sigma$ error. 
Up to $\ell \simeq 90$, both curves show a bias close to 1 within a $2 \sigma$ confidence interval. 
For $\ell \gtrsim 90$, non-linear effects become evident, with the observed spectra showing more power than expected from linear growth theory at small angular scales.}
\label{fig:Cls_linear_bias}
\end{figure}

\begin{figure}
    \begin{minipage}[b]{\linewidth}
        \centering
        \includegraphics[width=0.9\linewidth]{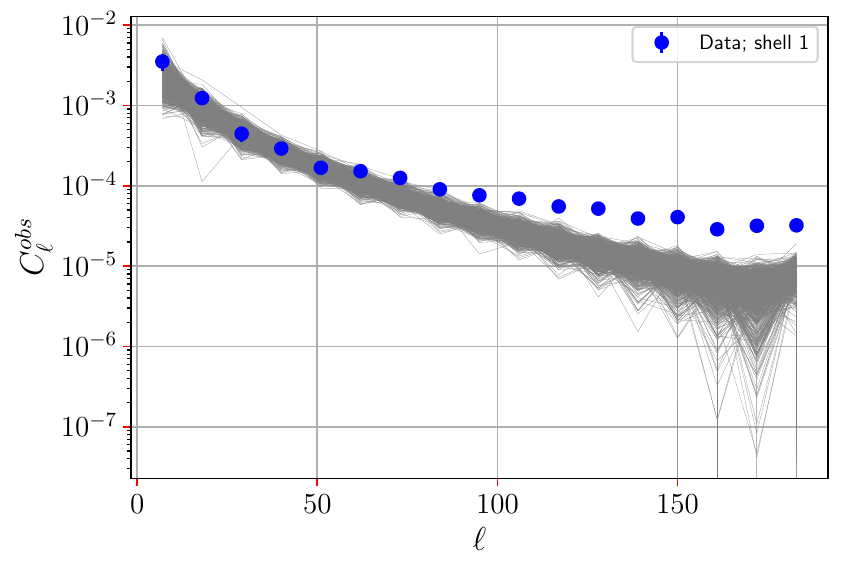}
        \includegraphics[width=0.9\linewidth]{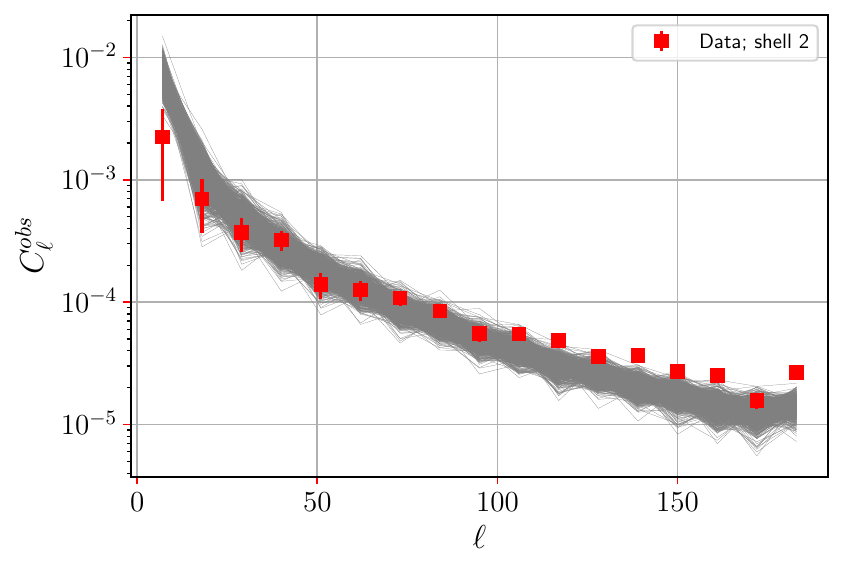}
\end{minipage}
\caption{
Comparative study of the angular power spectra calculated from the data, in each shell, with respect to the spectra obtained from the analyses of the set of log-normal simulations (mock data). 
{\bf Upper panel: Shell 1 analysis.} 
One observes that data and simulations agrees well until 
$\sim 2^{\circ}$, which corresponds to 
$\ell = 90$; for lower scales (i.e., larger $\ell$) there is an increasing difference 
between both data because the non-linear growth of structures dominates. 
{\bf Lower panel: Shell 2 analysis.} 
In this case, data and simulations agree well until $\sim 1^{\circ}$, corresponding to $\ell = 180$; 
notice that we are finding the same physical scale of non-linearity in both shells: 
$D_A^{\text{Shell}\,1} \times 2^{\circ} 
\simeq D_A^{\text{Shell}\,2} \times 1^{\circ}$, 
because the angular diameter distance to Shell 2 is, 
approximately, the double than the angular diameter distance 
to Shell 1.
}
\label{fig:Cls_mocks}
\end{figure}

To validate the results of our analyses, a comparison between observation and theory is necessary. 
In the $\Lambda$CDM model, within the theory of linear perturbations, the theoretical curve can be obtained using equation (\ref{eq:curva_teorica}). 
In the linear context, it is expected that the theoretical $C_{\ell}^{\text{th}}$ is proportional to the observed 
$C_{\ell}^{\text{obs}}$. 
Since $C_{\ell}^{\text{th}}$ is proportional to $b_0^2$, one can calculate the quantities
$\sqrt{C_{\ell}^{\text{obs}}/C_{\ell}^{\text{th}}}$, for $\ell \in [7,183]$. 
We present the results of this comparison in Figure~\ref{fig:Cls_linear_bias} for both shells. 
The shaded region represents the propagated error in $1\sigma$. Up to the multipole $\ell = 90$, both curves intersect, with a bias close to 1 within a $2\sigma$ confidence interval. However, as expected, for large multipoles 
the measurements show more power than predicted by linear theory. 
This could suggest that non-linear effects appear for scales $\ell \gtrsim 90$. 
Another key observation in this figure is that the curve for Shell 2 is closer to the linear theory than that of Shell 1, suggesting that our neighbourhood, $0 \leq z < 0.06$, has more developed structures that contribute to non-linear measurements. 
This is further illustrated in Figure~\ref{fig:Cls_mocks}, where the results are compared with the mock simulations. For the entire range of $\ell$, the data points for Shell 2 are in excellent agreement with the log-normal simulations. 
For Shell 1, a discrepancy is noticed for $\ell \gtrsim 90$. 
We cannot fail to mention that our mocks were generated with a non-linear power spectrum, but for very small scales 
the non-linear effects observed in our analyses 
contribute to the increase in power in a way not expected in the simulations~\citep{Mattewson22}.

\section{Conclusions and final remarks}
\label{sec:conclusions}

We employ the SDSS blue galaxies, excellent tracers of dark matter, to discover clustering properties of the network of cosmic structures, and for this we use various estimators that provide partial, but complementary, insights to solve the big cosmic puzzle. 

%

With the 2PACF, at small and large scales, we have studied quantitatively --through the $\beta$ parameter-- 
the clustering strengths of the SDSS blue galaxies finding them 
consistent with the clustering and growth of cosmic structures obtained from the analysis of the set of simulated mocks. 
The clustering is quantified by $\beta$, and the values obtained analysing the data make sense when compared with the average from a similar analysis made with the mocks; 
instead the growth of cosmic structures is confirmed by comparison between the $\beta$ values in both shells. 
In fact, the confront of $\beta$ between different epochs reflects the clustering process due to gravitational instability shaping the matter distribution, with Shell 1 showing more evolved and clustered structures compared to Shell 2, which maps a (slightly) younger epoch of the Universe. 
However, with the 2PACF at large scales we obtain even more information, since one can recognize the presence of large over-dense and/or under-dense cosmic structures in the Areas in study (see Sections~\ref{2pacf-LA} and~\ref{sec:robustness}). 
These results enhance our understanding of the 
evolutionary processes that have shaped the large-scale structure of the Universe across cosmic time~\citep{Ando18,Fang20,Avila2021}.

Our results with the 2PACF serve as a consistency test of the model, represented in the mocks, and at the same time they represent a directional analysis of the cosmic tracer displayed in the selected 12 Areas, confirming the validity of the statistical isotropy in the Local Universe in each shell.

In our analyses of the angular power spectrum estimator, we inspect the two redshift shells 
and compare the results 
in the context of the $\Lambda$CDM model, comparing observational results with theoretical predictions. For multipoles 
$\ell \lesssim 90$, the observed power spectra for both shells show a bias close to 1, consistent with linear theory within a $2\sigma$ confidence interval. 
However, deviations from linear theory are observed for large multipoles, particularly for Shell 1, which corresponds to the Local Universe. This suggests that local structures contribute significantly to non-linearities, with a notable increase in power at smaller scales. The comparison with log-normal simulations further confirms these effects, especially for Shell 1, where the disagreement between data and simulations is observed for the lower scales, that is, $\ell \gtrsim 90$. 

Moreover, we also take advantage of the CDF tool which provides useful insights to the third dimension, the radial distances 
(they are calculated using cosmography and are given in units Mpc $h^{-1}$, where 
$H_0 \equiv 100 \,h$ km s$^{-1}$ Mpc$^{-1}$). 
By comparing the CDF of a set of Areas both in 
Shell~1 and Shell~2, distinctive patterns emerge, namely regions revealing strong matter clustering followed by voids, a signature consistent across the Areas within the same shell, but showing a different pattern when comparing the same Area across the two shells. 
This, once more, reinforces the evolutionary differences, as Shell 1 corresponds to older structures than Shell 2. 
Moreover, the CDF analysis reveals that these features, including the presence of voids and clusters, are independent of the number of galaxies in each Area. 
These findings complement our angular, or 2D, clustering studies.

Additionally, the maps presented in Figure~\ref{fig:maps_shells_64} further illustrate the evolutionary disparity between the two redshift shells. While Shell 1 exhibits a more developed network of cosmic structures, Shell 2 reflects a younger epoch, where these cosmic structures are still in the process of formation. This strengthens the results observed in the angular and radial analyses, emphasizing the dynamic of structure formation. 
Furthermore, wedge plots complement the information provided by CDF, giving 
light to understand the imprints left by cosmic structures in the results 
obtained with other tools, in summary, helping to unveil features partially 
or totally hidden in other examinations.

Finally, the number-count maps, displayed in 
Figure~\ref{fig:maps_shells_64}, illustrates well our conclusions, summarizing the results of our study. 
We performed analyses that were consistent with the $\Lambda$CDM model, separately, in each shell. 
But this consistency works just as a snapshot in each shell, however the $\Lambda$CDM model describes also the dynamics of the matter clustering evolution, and therefore it can also be tested in this regard. 
A comparison between shells of the clustering features, done with diverse estimators, should make such cosmic evolution evident. 
This is precisely what has been done in subsections~\ref{2pacf-SA},~\ref{2pacf-LA}, and~\ref{sec:Cl_results}, and is well illustrated in Figure~\ref{fig:maps_shells_64}: the older part of the Universe seen in Shell~1, $z \simeq 0$, 
shows structures suggestive of filament-like morphology, where various large over-densities are present but also well-defined void structures are noticeable, 
instead such structures are not remarkable in the younger part of the Universe represented in Shell~2. 


\section*{Acknowledgements}
C.F. and A.B. thank the Coordenação de Aperfeiçoamento de Pessoal de Nível Superior (CAPES) and Conselho Nacional de Desenvolvimento Científico e Tecnológico (CNPq) for their grants under which this work was carried out. 
F.A. thanks to Funda\c{c}\~{a}o Carlos Chagas Filho de Amparo \`{a} Pesquisa do Estado do Rio de Janeiro (FAPERJ), Processo SEI-260003/001221/2025, for the financial support.
This work was carried out using computational resources provided by the Data Processing Center of the National Observatory (CPDON).

\section*{Data Availability}
The datasets underlying this article are available in Zenodo
at doi: \href{https://zenodo.org/records/17144753}{10.5281/zenodo.17144753}.

\appendix

\section{Comparison between the distribution of \texorpdfstring{$\beta$}{} values from mocks and SDSS data}
\label{app:percentiles}

\begin{figure*}[!htpb]
    \centering
    \includegraphics[width=\linewidth]{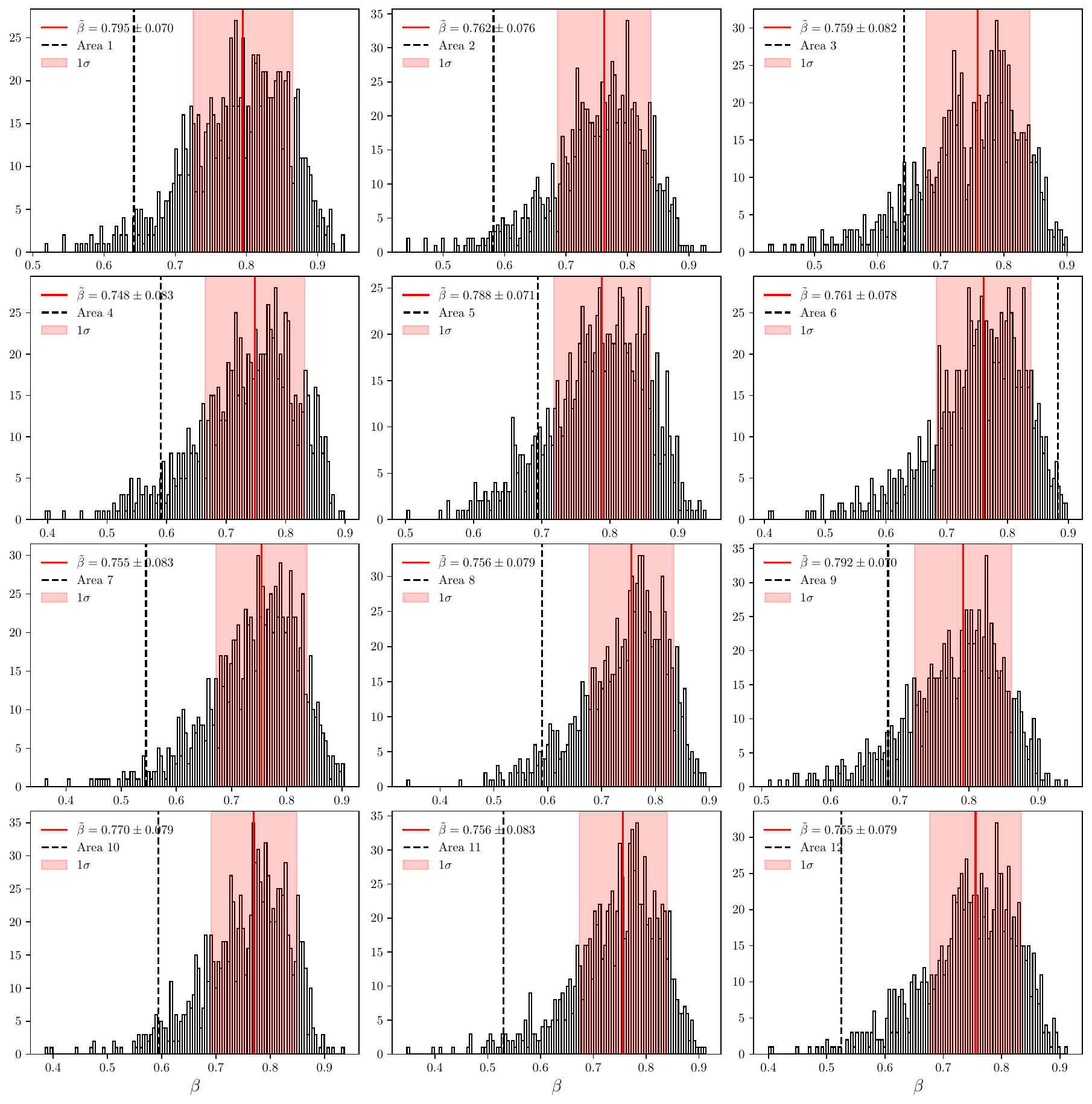}
    \caption{Distribution of the best-fit $\beta$ values from $1,000$ Area-mocks in Shell 1 (small scales). The black histogram shows the mock distribution, the red solid line is the median value, and the red shaded region represents the $1\sigma$ interval. Dashed vertical lines indicate the observational $\beta$ values obtained from SDSS in the $12$ sky regions.}
    \label{fig:hist_shell1_small}
\end{figure*}

\begin{figure*}[!htpb]
    \centering
    \includegraphics[width=\linewidth]{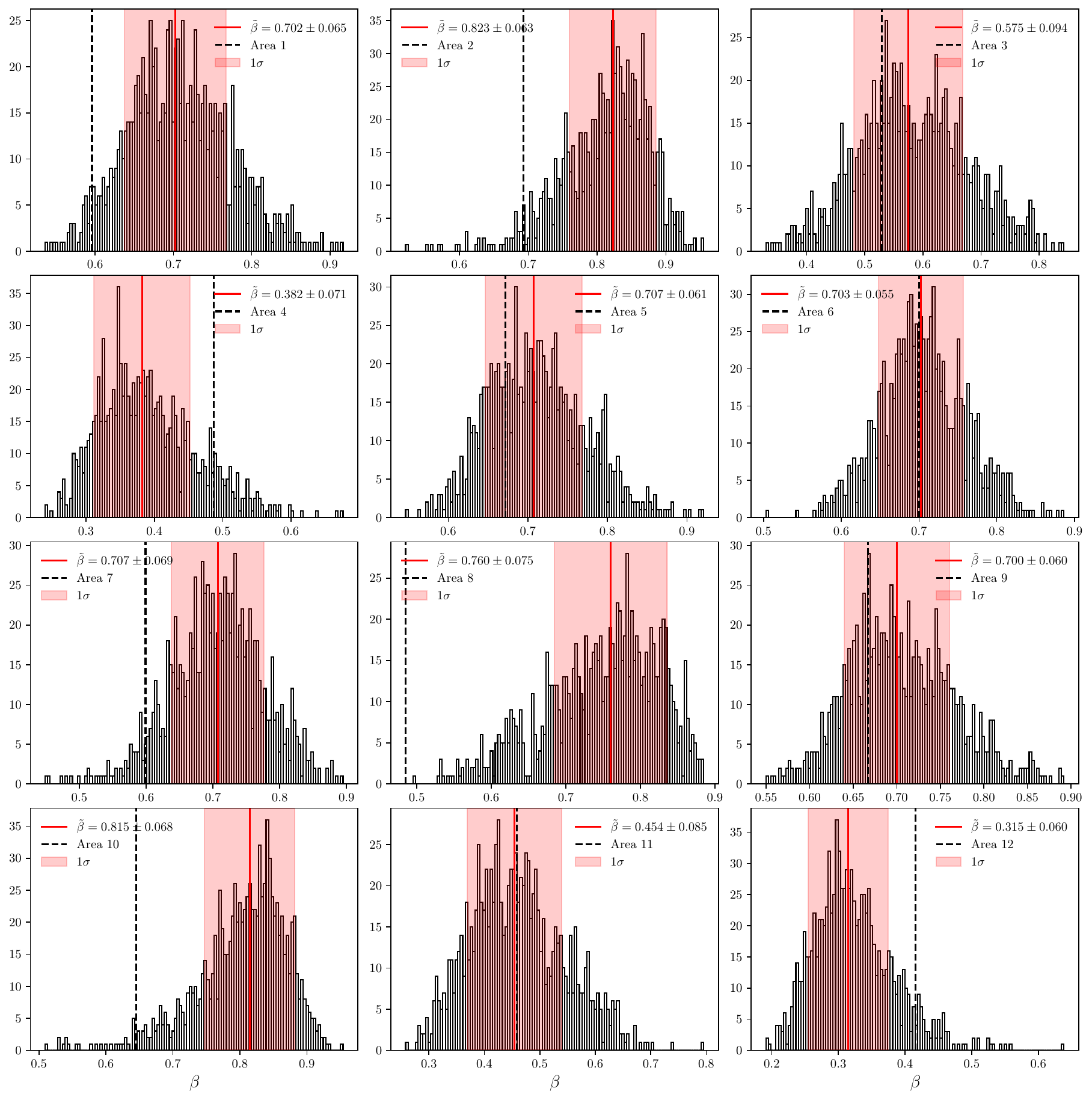}
    \caption{Same as Figure~\ref{fig:hist_shell1_small}, but for Shell 2 (small scales).}
    \label{fig:hist_shell2_small}
\end{figure*}

\begin{figure*}[!htpb]
    \centering
    \includegraphics[width=\linewidth]{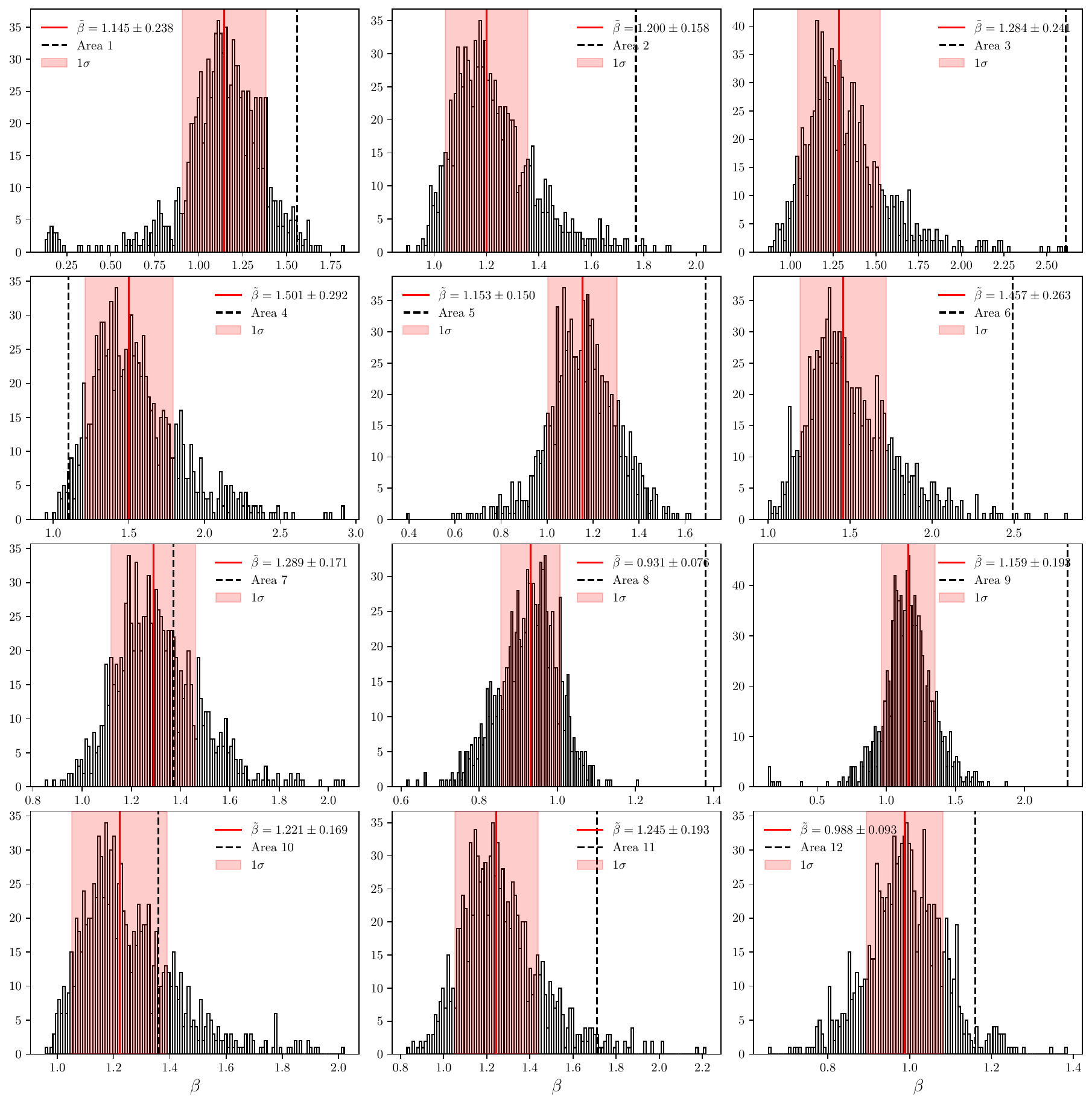}
    \caption{Same as Figure~\ref{fig:hist_shell2_small}, but for Shell 1 (large scales).}
    \label{fig:hist_shell1_large}
\end{figure*}

\begin{figure*}[!htpb]
    \centering
    \includegraphics[width=\linewidth]{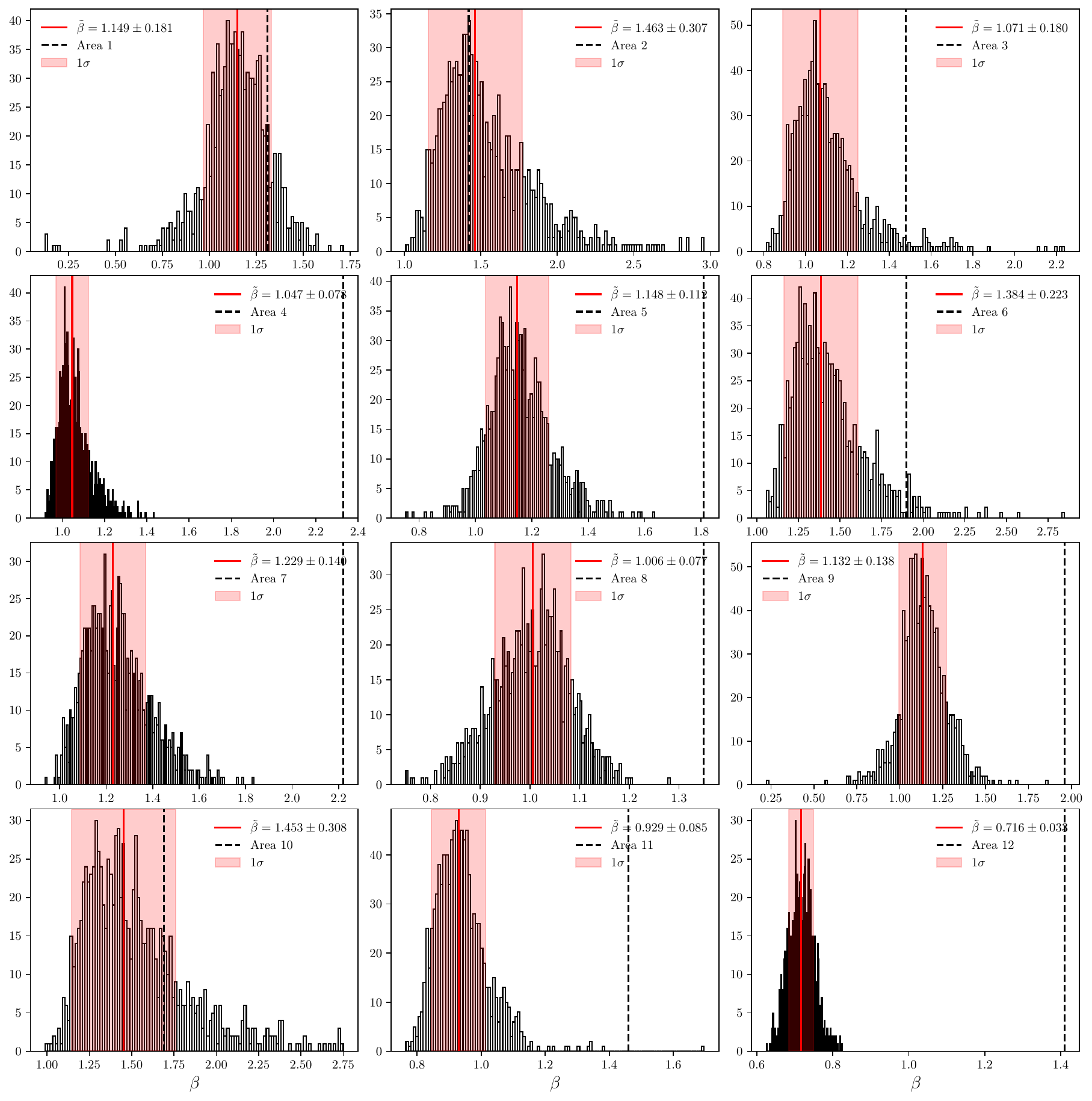}
    \caption{Same as Figure~\ref{fig:hist_shell1_large}, but for Shell 2 (large scales).}
    \label{fig:hist_shell2_large}
\end{figure*}

To complement the discussion of Section~\ref{sec:results}, we present the distribution of the best-fit $\beta$ values obtained from the mock catalogs compared with those measured from the SDSS data. For each of the $12$ sky regions in both redshift shells, we generated $1,000$ Area-mocks and obtained the corresponding distributions. The results are shown in Figures~\ref{fig:hist_shell1_small}--\ref{fig:hist_shell2_large}.

In these figures, the black histograms display the $\beta$ values obtained from the mocks, while the solid red line and shaded region indicate the median and $1\sigma$ interval, respectively. 
The observational value in the corresponding Area is shown as a dashed line.

As discussed in Section~\ref{sec:results}, the comparison highlights distinct behaviors across shells. On small scales, the $\beta$ values 
from the SDSS blue galaxies tend to fall below the expected values from the mocks, 
whereas in the large scales they fall consistently above the mock medians.

Thus, these figures provide a complement to the results summarized in Table~\ref{tab:percentiles} and presented in Figures~\ref{fig:beta-distribution_small}~and~\ref{fig:beta-distribution_large}.

\section{Robustness of Angular Power Spectrum Results with Varying Resolution}
\label{app:healpy_32}

\begin{figure}
\begin{minipage}[b]{\linewidth}
\centering
\includegraphics[width=0.9\linewidth]{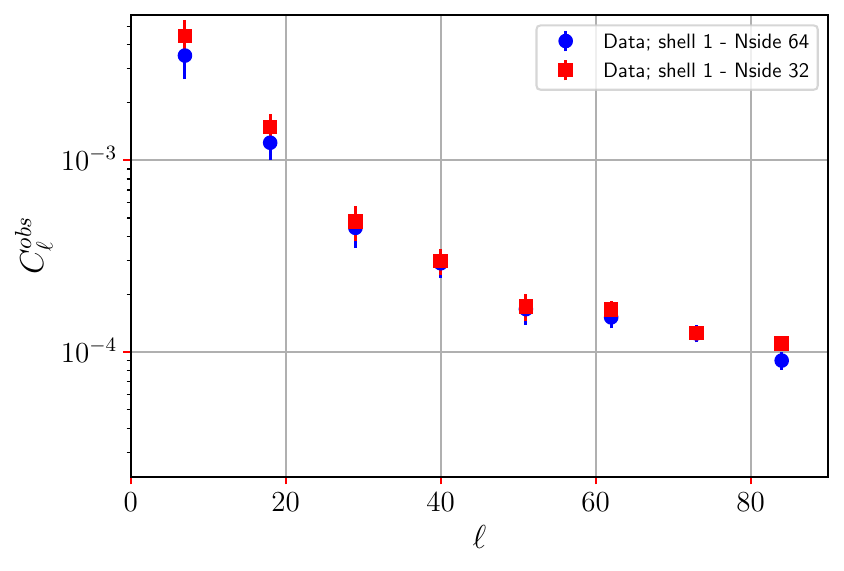}
\includegraphics[width=0.9\linewidth]{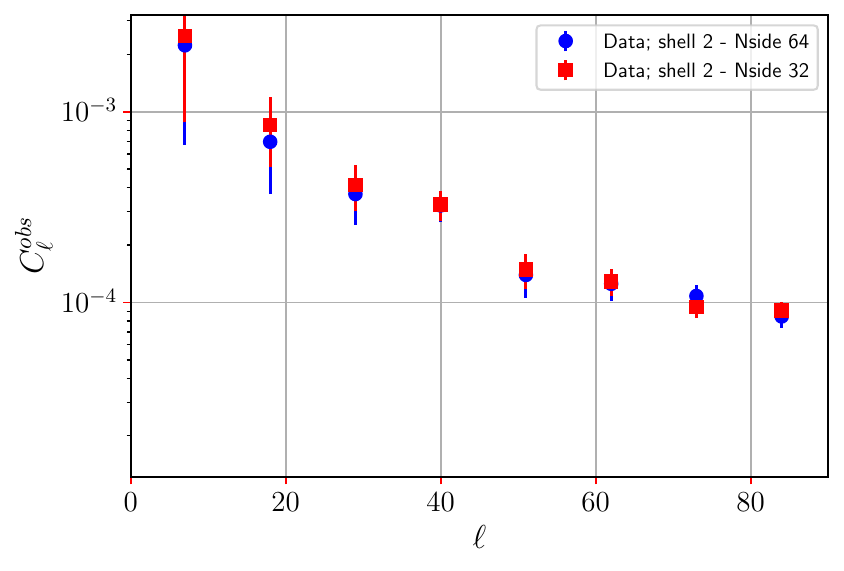}
\end{minipage}
\caption{
Consistency test for different angular resolutions. 
We compute the angular power spectra, $C_{\ell}$, from data in the two shells to verify the consistency of our results when considering two angular resolutions, namely, $N_{\text{side}}=32$ and $N_{\text{side}}=64$. 
In both plots, our results are presented up to the common multipole $\ell=84$, due to their different limit resolutions. 
As observed, there is no significant difference in the results obtained in both cases.
}
\label{fig:Cls_nside_32_64}
\end{figure}

To assess the robustness of the result obtained with $N_{\text{side}}=64$, we present the angular power spectrum for $N_{\text{side}}=32$. 
Figure~\ref{fig:Cls_nside_32_64} compares the $C_{\ell}$ measurements for both shells between $N_{\text{side}}=32$ and $N_{\text{side}}=64$. 
Note that the results are shown up to $\ell=84$ due to loss of resolution in the pixels with the decrease in $N_{\text{side}}$. 
Overall, there are no significant variations in the $C_{\ell}$ measurements 
for both shells when reducing the map resolution, which strengthens our 
results and their interpretation presented in 
Section~\ref{sec:Cl_results}.

\bibliographystyle{aasjournalv7}
\bibliography{bib}

\end{document}